\documentclass[12pt]{amsart}
\usepackage[utf8]{inputenc}
\usepackage[english]{babel}
\usepackage{amsmath, amssymb, amsthm, amscd,color,comment}
\usepackage{cancel}
\usepackage{cite}
\usepackage{alltt}
\usepackage[dvipsnames]{xcolor}
\usepackage{array}
\usepackage[small,bf,labelsep=period]{caption}
\usepackage{mathtools}
\usepackage{hyperref}

\usepackage{enumerate}
\newtheorem{theorem}{Theorem}[section]
\newtheorem{definition}[theorem]{Definition}
\newtheorem{example}[theorem]{Example}

\newtheorem{remark}[theorem]{Remark}
\newtheorem{lemma}[theorem]{Lemma}
\newtheorem{corollary}[theorem]{Corollary}
\newtheorem{proposition}[theorem]{Proposition}

\newcommand{\F}{\mathbb{F}}

\DeclareMathOperator{\charac}{char}
\DeclareMathOperator{\lpf}{lpf}

\def\F{\mathbf F}

\def\dim{{\rm dim}}
\def\deg{{\rm deg}}

\def\neg1{\text{\boldmath$1$}}

\def\neg1{\text{\boldmath$1$}}

\def\negc{\text{\boldmath$c$}}
\def\negx{\text{\boldmath$x$}}
\def\nega{\text{\boldmath$a$}}

\begin{document}

\title[On cyclic codes generated by cyclotomic polynomials]{On cyclic codes generated by cyclotomic polynomials}



\thanks{The first author was partially supported by CNPq Grant 303983/2026-4 and FAPEMIG Grant APQ-06634-24, RED-0013-21. The second author was partially supported by FAPEMIG Grant 2022/16369-2. The third author was partially funded by CNPq Grant 306270/2023-4 and FAPEMIG Grant APQ-01443-23.}

\author{Alonso S. Castellanos, Gyan Morales Solis, and Guilherme Tizziotti}

\address{Instituto de Matem\'{a}tica e Estat\'{i}stica, Universidade Federal de Uberl\^{a}ndia, Campus Santa M\^{o}nica, CEP 38400-902, Uberl\^{a}ndia, Brazil}
\email{alonso.castellanos@ufu.br}

\address{Instituto de Matem\'{a}tica e Estat\'{i}stica, Universidade Federal de Uberl\^{a}ndia, Campus Santa M\^{o}nica, CEP 38400-902, Uberl\^{a}ndia, Brazil}
\email{gyan.solis@ufu.br}

\address{Instituto de Matem\'{a}tica e Estat\'{i}stica, Universidade Federal de Uberl\^{a}ndia, Campus Santa M\^{o}nica, CEP 38400-902, Uberl\^{a}ndia, Brazil}
\email{guilhermect@ufu.br}

\begin{abstract}
Let $n$ be a composite positive integer coprime to $\operatorname{char}(\mathbb{F}_q)$
and let $t = \operatorname{lpf}(n)$ be the least prime factor of $n$.
In this paper we study the cyclic codes $\mathcal{C}_{n,t} = \langle Q_n(X)Q_t(X)\rangle$
and $\mathcal{C}_{n,t,1} = \langle Q_n(X)Q_t(X)Q_1(X)\rangle$ over $\mathbb{F}_q$,
where $Q_d(X)$ denotes the $d$-th cyclotomic polynomial. These codes extend the
families $\mathcal{C}_n = \langle Q_n(X)\rangle$ and
$\mathcal{C}_{n,1} = \langle Q_n(X)Q_1(X)\rangle$ recently studied by Bhagat and
Sarma. We determine the exact minimum distances of $\mathcal{C}_{n,t}$ and
$\mathcal{C}_{n,t,1}$, showing that $d(\mathcal{C}_{n,t,1}) = 2t$ and that
$d(\mathcal{C}_{n,t})$ equals $t$ if $t^2 \nmid n$, and $2t$ if $t^2 \mid n$.
For the dual codes, we establish $d(\mathcal{C}_{n,t}^\perp) = 2^{\omega(n)}$,
handling the case $\omega(n) \le 2$ via the Hartmann--Tzeng bound, and the case
$\omega(n) \ge 3$ via a direct-sum decomposition
$\mathcal{C}_{n,t}^\perp = \mathcal{C}_n^\perp \oplus \mathcal{V}_t$ and the
tensor-product structure of $\mathcal{C}_n^\perp$ over its prime-power components.
As applications, we show that all four code families are linear complementary dual
(LCD) and locally recoverable (LRC), and we determine their locality parameters,
identifying several infinite families that are $r$-optimal or $d$-optimal with
respect to the Singleton-type bound for LRC codes.
\end{abstract}

\maketitle

\section*{Introduction}

For a prime power $q$ and a positive integer $n$ with $\gcd(n, q) = 1$, cyclic
codes of length $n$ over $\mathbb{F}_q$ are ideals of the quotient ring
$\mathbb{F}_q[X]/\langle X^n - 1\rangle$, and are generated by divisors of
$X^n - 1$. Since $X^n - 1 = \prod_{d \mid n} Q_d(X)$ over $\mathbb{F}_q$,
where $Q_d(X)$ denotes the $d$-th cyclotomic polynomial
(see~\cite{LidlNiederreiter1994}), cyclic codes generated by products of cyclotomic
polynomials form a natural and algebraically well-structured subclass. Well-known
families such as BCH codes and Reed--Solomon codes fall within this framework; we
refer the reader to~\cite{HuffmanPless2003, MacWilliamsSloane1977, LingXing2004}
for background.

For a positive integer $n > 1$ coprime to $\operatorname{char}(\mathbb{F}_q)$, the
cyclic codes generated by $Q_n(X)$ and by $Q_n(X)Q_1(X)$ were studied by Bhagat
and Sarma in~\cite{BhagatSarma2025, BhagatSarma2026v2}. Denoting these codes by
$\mathcal{C}_n = \langle Q_n(X)\rangle$ and
$\mathcal{C}_{n,1} = \langle Q_n(X)Q_1(X)\rangle$, and letting $\omega(n)$ the number of distinct prime factors of $n$ and $\operatorname{lpf}(n)$ the least prime factor of $n$, they determined the exact
parameters
\[
\mathcal{C}_n \colon [n,\, n - \varphi(n),\, \operatorname{lpf}(n)]_q, \qquad
\mathcal{C}_{n,1} \colon [n,\, n - \varphi(n) - 1,\, 2\operatorname{lpf}(n)]_q,
\]
as well as the parameters $[n, \varphi(n), 2^{\omega(n)}]_q$ and
$[n, \varphi(n) + 1, 2^{\omega(n)}]_q$ of their Euclidean duals, and established
that all four codes are LCD-LRC codes with explicit locality parameters and
optimality properties.

In this paper, we extend this study to two families obtained by incorporating an
additional cyclotomic factor. Let $n$ be a composite integer coprime to
$\operatorname{char}(\mathbb{F}_q)$ and let $t = \operatorname{lpf}(n)$ denote its
least prime factor. We consider the cyclic codes
\[
\mathcal{C}_{n,t} = \langle Q_n(X)Q_t(X)\rangle
\quad \text{and} \quad
\mathcal{C}_{n,t,1} = \langle Q_n(X)Q_t(X)Q_1(X)\rangle.
\]
The inclusions $\mathcal{C}_{n,t} \subsetneq \mathcal{C}_n$ and
$\mathcal{C}_{n,t,1} \subsetneq \mathcal{C}_{n,1}$ imply
$\mathcal{C}_n^\perp \subsetneq \mathcal{C}_{n,t}^\perp$ and
$\mathcal{C}_{n,1}^\perp \subsetneq \mathcal{C}_{n,t,1}^\perp$, so the
codes $\mathcal{C}_{n,t}$ and $\mathcal{C}_{n,t,1}$ have smaller dimension
than their counterparts in~\cite{BhagatSarma2025}, while their duals have larger dimension and
minimum distance at most $2^{\omega(n)}$.

We determine the exact minimum distances of $\mathcal{C}_{n,t}$ and
$\mathcal{C}_{n,t,1}$ in Section~\ref{sec:dist_cnt}. We show that
$d(\mathcal{C}_{n,t,1}) = 2t$ in all cases, while $d(\mathcal{C}_{n,t})$ exhibits
a dichotomy depending on the divisibility of $n$ by $t^2$: it equals $t$ if
$t^2 \nmid n$, and $2t$ if $t^2 \mid n$. The upper bounds are achieved by explicit
low-weight codewords, and the lower bounds follow either from inclusion inequalities
or from the BCH bound applied to the defining set of $\mathcal{C}_{n,t}$.

For the dual codes, we establish $d(\mathcal{C}_{n,t}^\perp) = 2^{\omega(n)}$
(assuming $n > t^2$). The case $\omega(n) \le 2$ is treated in
Section~\ref{sec:dual_omega_le2} via the Hartmann--Tzeng bound~\cite{HartmannTzeng1972},
together with direct arguments for the special cases $n = \ell^a$ and $n = p_1p_2$.
The case $\omega(n) \ge 3$, handled in Section~\ref{sec:dual_omega_ge3}, requires
a more structural approach. We prove that $\mathcal{C}_{n,t}^\perp$ decomposes as
the direct sum $\mathcal{C}_n^\perp \oplus \mathcal{V}_t$, where $\mathcal{V}_t$
is a periodic cyclic subspace of $\mathbb{F}_q^n$. We then use the
tensor-product factorization
\[
\mathcal{C}_n^\perp = \Psi_k\bigl(\mathcal{C}_{p_1^{a_1}}^\perp \otimes \cdots
\otimes \mathcal{C}_{p_k^{a_k}}^\perp\bigr),
\]
established via the Chinese Remainder Theorem as a generalization of a result
of~\cite{BhagatSarma2026v2}, together with parity constraints along each tensor
direction and the periodic structure of $\mathcal{V}_t$, to prove that every
nonzero codeword of $\mathcal{C}_{n,t}^\perp$ has weight at least $2^{\omega(n)}$.

Section~\ref{sec:lcd_lrc} presents two applications of these results.
First, we show that $\mathcal{C}_{n,t}$, $\mathcal{C}_{n,t,1}$, and their duals
are \textit{linear complementary dual} (LCD) codes, a class introduced by
Massey~\cite{Massey1992} and of renewed interest due to applications in
side-channel attack countermeasures in cryptographic hardware
implementations~\cite{CarletGuilley2016}. The LCD property follows from the fact
that the generator polynomials of all four codes are products of cyclotomic
polynomials, which are self-reciprocal for $d > 1$. Second, we show that all four
families are \textit{locally recoverable} (LRC) codes in the sense
of~\cite{GopalanHuangSimitciYekhanin2012}, by relating the locality of each cyclic
code to the minimum distance of its dual code. In particular, we identify several
infinite families that are $r$-optimal or simultaneously $r$-optimal and $d$-optimal
with respect to the Singleton-type bound for LRC codes established
in \cite{GopalanHuangSimitciYekhanin2012} (see also~\cite{BhagatSarma2025}).

The paper is organized as follows. Section~\ref{sec:prelim} collects the necessary
background on cyclotomic polynomials, cyclic codes, and LCD and LRC codes.
Section~\ref{sec:dist_cnt} determines the minimum distances of $\mathcal{C}_{n,t}$
and $\mathcal{C}_{n,t,1}$. Section~\ref{sec:dual_omega_le2} treats
$d(\mathcal{C}_{n,t}^\perp)$ for $\omega(n) \le 2$, and
Section~\ref{sec:dual_omega_ge3} handles the case $\omega(n) \ge 3$.
Section~\ref{sec:lcd_lrc} presents the LCD and LRC applications.

\section{Preliminaries and Notation}
\label{sec:prelim}

Let $p$ be a prime integer, $q$ be a power of $p$, and $\mathbb{F}_q$ be the finite field with $q$ elements. We denote the characteristic of $\mathbb{F}_q$ by $\operatorname{char}(\mathbb{F}_q)$. Let $\mathbb{N}$ denote the set of non-negative integers. For $0 \neq n \in \mathbb{N}$, we denote by $\omega(n)$ the number of distinct prime factors of $n$, and by $\operatorname{lpf}(n)$ the least prime factor of $n$. If an integer $a$ divides the integer $b$ we write $a \mid b$. Otherwise we write $a \nmid b$.

\subsection{Cyclotomic polynomials.} Let $n$ be a positive integer not divisible by $p$ and let $\alpha$ be a primitive $n$-th root of unity over $\mathbb{F}_q$. The $n$-th \textit{cyclotomic polynomial over} $\mathbb{F}_q$, denoted by $Q_n(X)$, is defined as the monic polynomial whose roots are precisely all primitive $n$-th roots of unity, that is:
	\begin{equation} \label{n cyclotomic polynomial}
			Q_n(X) := \prod_{\substack{1 \le k \le n \\ \gcd(k,n)=1}} (X - \alpha^k).
		\end{equation}
Note that, the polynomial $Q_n(X)$ is independent of the choice of $\alpha$ and its degree is given by Euler's totient function $\varphi$, $\deg(Q_n(X)) = \varphi(n)$. In addition, $Q_n(X) \in \mathbb{F}_q[X]$ and we have the following fundamental identity decomposing $X^n - 1$ into the product of the cyclotomic polynomials associated with the divisors of $n$.

	\begin{theorem}\cite[Theorem 2.45]{LidlNiederreiter1994}
		\label{thm:descomposicao_xn-1}
		Let $n \in \mathbb{N}$ be such that $\gcd(n, p) = 1$. Then,
		\begin{equation*}
			X^n - 1 = \prod_{d \mid n} Q_d(X).
		\end{equation*}

	\end{theorem}

The following property of the roots of cyclotomic polynomials of prime index will be useful in this work.

	\begin{remark} \label{lema:raizes_Qt_primo}
		Let $t$ be a prime such that $\operatorname{char}(\mathbb{F}_q) \nmid t$. By Theorem \ref{thm:descomposicao_xn-1} we have that $X^{t} - 1 = Q_1(X)Q_t(X) $. So, $Q_t(X) = 1 + X + \dots + X^{t-1}$, since $Q_1(X) = X-1$. Moreover, if $\zeta$ is a primitive $n$-th root of unity and $t$ divides $n$, then $\eta = \zeta^{n/t}$ has order $\frac{n}{\gcd(n, n/t)} = t$, and so $\eta$ is a primitive $t$-th root of unity. Therefore, we can conclude that $Q_t(\zeta^{n/t}) = 0$.
		\end{remark}

Let $m$ be a positive integer divisible by a prime $\ell \neq \operatorname{char}(\mathbb{F}_q)$. Then $Q_{\ell m}(X) = Q_m(X^{\ell})$ (see \cite[Exercise 2.57]{LidlNiederreiter1994}). As a consequence of this fact we get the following result.

\begin{lemma}
		\label{lema:paridade_Qn}
		Let $n$ be a positive integer divisible by $4$ and such that $\operatorname{char}(\mathbb{F}_q) \nmid n$. Then, $Q_n(X) = Q_{n/2}(X^2)$. In particular, $Q_n(-X) = Q_n(X)$.
	\end{lemma}

\begin{lemma}
	\label{teo:divisibilidade_P_Qt}
	Let $n$ be a composite integer with $\operatorname{char}(\mathbb{F}_q) \nmid n$ and let $t = \operatorname{lpf}(n)$ with $n > t^2$. Let $P(X) = \frac{X^n-1}{X^s-1}$, where $s=\frac{n}{t}$. Then the cyclotomic polynomial $Q_t(X)$ divides $P(X)$ if and only if $t^2 \nmid n$.
\end{lemma}

\begin{proof}
Let $\zeta$ be a primitive $n$-th root of unity over $\mathbb{F}_q$. The roots of $P(X)$ are the roots of $X^n - 1$ that are not roots of $X^s - 1$. A power $\zeta^j$ is a root of $X^s - 1$ if and only if $n \mid js$. Since $n = st$, the condition $n \mid js$ is equivalent to $t \mid j$. Thus, the set of roots of $P(X)$ is given by those $\zeta^j$ such that $t \nmid j$. The roots of $Q_t(X)$ are the primitive $t$-th roots of unity, given by $\zeta^{ks}$ with $\gcd(k, t) = 1$.

Since $\operatorname{char}(\mathbb{F}_q) \nmid n$, the polynomial $X^n-1$ has only simple roots. Therefore, $Q_t(X)$ divides $P(X)$ if and only if all roots of $Q_t(X)$ are also roots of $P(X)$. For this to occur, the exponent $ks$ must not be a multiple of $t$, that is, $t \nmid ks$.

	Now, $k$ and $t$ are coprime, and so the condition $t \nmid ks$ is equivalent to $t \nmid s$. Setting $s = n/t$, we find that $t \nmid (n/t)$, and thus $t^2 \nmid n$.

	Therefore, if $t^2 \mid n$, the roots of $Q_t(X)$ are zeros of $X^s-1$. Then $\gcd(Q_t(X), P(X)) = 1$ and $Q_t(X)$ does not divide $P(X)$. Conversely, if $t^2 \nmid n$, then $Q_t(X)$ divides $P(X)$.
\end{proof}

For $f(X) \in \mathbb{F}_q[X]$ with $\deg f(X)=\lambda$, the polynomial $f^*(X) := X^{\lambda} f(X^{-1})$ is called the \textit{reciprocal polynomial} of $f(X)$. If $f(X) = f^*(X)$, we say that $f(X)$ is a \textit{self-reciprocal polynomial}.

For $n > 1$, the cyclotomic polynomial $Q_n(X)$ is a self-reciprocal polynomial. Indeed, the set of primitive $n$-th roots of unity is closed under $\alpha \mapsto \alpha^{-1}$, so $Q_n^*(X) = Q_n(0)\,Q_n(X)$, and $Q_n(0) = 1$: for $n > 2$ the roots can be paired as $\{\alpha, \alpha^{-1}\}$ with $\alpha\alpha^{-1}=1$ and $\varphi(n)$ is even, while $Q_2(0) = 1$. On the other hand, $Q_1^*(X) = -Q_1(X)$.

\subsection{Cyclic codes.} A \textit{linear code} $\mathcal{C}$ of length $n$ and dimension $k$ over $\mathbb{F}_q$, that we simply refer to as $[n,k]_q$-code, is a $k$-dimensional subspace of the vector space $\mathbb{F}_{q}^{n}$. A \textit{generator matrix} $G$ for a $[n,k]_q$-code $\mathcal{C}$ is a $k$ by $n$ matrix for which the rows are a basis of $C$. If $G$ is a generator matrix for $\mathcal{C}$, then $\mathcal{C} = \{ \nega G \mbox{ ; } \nega \in \mathbb{F}_{q}^k{} \}$. The \textit{weight} of an element $\negc = (c_0, \ldots , c_{n-1})\in \mathcal{C}$ is defined by $\operatorname{wt}(\negc):= \# \{ 0 \leq i \leq n-1 \mbox{ ; } c_i \neq 0 \}$ and the \textit{minimum distance} of $\mathcal{C}$ is given by $d(\mathcal{C}) = \min \{\operatorname{wt}(\negc) \mbox{ ; } \negc \in \mathcal{C} \setminus \{\mathbf{0}\}\}$. An $[n,k]_q$-code with minimum distance $d$ is called a $[n,k,d]_q$-code.

Two linear codes $\mathcal{C}_1$ and $\mathcal{C}_2$ of length $n$ over $\mathbb{F}_q$, where $G_1$ is a generator matrix for $\mathcal{C}_1$, are called \textit{permutation equivalent} if there exists a $n \times n$ permutation matrix $P$ such that $G_1P$ is a generator matrix for $\mathcal{C}_2$.

An $[n,k,d]_q$-code is called \textit{cyclic} if for every $\negc = (c_0, c_1, \ldots , c_{n-1}) \in \mathcal{C}$, the vector $(c_{n-1}, c_0 , \ldots , c_{n-2})$ also belongs to $\mathcal{C}$. A cyclic $[n,k,d]_q$-code $\mathcal{C}$ can be identified with an ideal of the quotient ring $R_{q,n} := \mathbb{F}_q[X]/\langle X^n - 1 \rangle$ by associating each vector $(c_0, c_1, \ldots , c_{n-1}) \in \mathbb{F}_{q}^{n}$ with the residue class
$$
\negc (X) + \langle X^n - 1 \rangle \in R_{q,n},
$$
where $\negc (X) := c_0 + c_1 X + \ldots + c_{n-1}X^{n-1}$. We denote this residue class by $\overline{\negc}(X)$

 With this identification and using that $R_{q,n}$ is a principal ideal ring, the cyclic $[n,k,d]_q$-code $\mathcal{C}$ is generated by a unique monic polynomial $g_\mathcal{C}(X)$ dividing $X^n - 1$, that is, $\mathcal{C} = \langle g_\mathcal{C}(X) \rangle$. It is well known in the theory of error-correcting codes that $k = n - \deg(g_\mathcal{C}(X))$. The polynomial $h_{\mathcal{C}}(X) := \frac{X^n - 1}{g_{\mathcal{C}}(X)}$ is called \textit{parity-check polynomial} of $\mathcal{C}$ and $\deg(h_{\mathcal{C}}(X)) = k$.

The \textit{dual} of an $[n,k]_q$-code $\mathcal{C}$ over $\mathbb{F}_q$ is the linear code defined by $\mathcal{C}^{\perp} := \{ \negx \in \mathbb{F}_{q}^{n} \mbox{ ; } \langle \negx , \negc \rangle = 0 \mbox{, for all } \negc \in \mathcal{C} \}$, where $\langle  ,  \rangle$ denotes the Euclidean inner product on $\mathbb{F}_{q}^{n}$. If $\mathcal{C}$ is a cyclic $[n,k]_q$-code, then $\mathcal{C}^\perp$ is a cyclic $[n,n-k]_q$-code and $\mathcal{C}^\perp = \langle h_{\mathcal{C}}^*(X)/h_{\mathcal{C}}(0) \rangle$, where $h_{\mathcal{C}}^*$ is the reciprocal polynomial of parity-check polynomial $h_{\mathcal{C}}(X)$.

		We now introduce two auxiliary lemmas on the duals of cyclic codes whose generator polynomials are products of cyclotomic polynomials. They will be used repeatedly in Sections~3 and~4.

		\begin{lemma}\label{lem:dual_ciclotomico}
Let $n$ be a positive integer with $\operatorname{char}(\mathbb{F}_q) \nmid n$, let $S$ be a set of divisors of $n$, and let $\mathcal{C}$ be the cyclic code of length $n$ over $\mathbb{F}_q$ with generator polynomial $g_{\mathcal{C}}(X) = \prod_{d \in S} Q_d(X)$. Then the dual code $\mathcal{C}^\perp$ is cyclic with generator polynomial
\[
g_{\mathcal{C}^\perp}(X) = \frac{X^n - 1}{g_{\mathcal{C}}(X)} = \prod_{\substack{d \mid n \\ d \notin S}} Q_d(X).
\]
\end{lemma}

\begin{proof}
By Theorem \ref{thm:descomposicao_xn-1}, $X^n - 1 = \prod_{d \mid n} Q_d(X)$, so $g_{\mathcal{C}}(X)$ divides $X^n - 1$ and the parity-check polynomial of $\mathcal{C}$ is
\[
h_{\mathcal{C}}(X) = \frac{X^n - 1}{g_{\mathcal{C}}(X)} = \prod_{\substack{d \mid n \\ d \notin S}} Q_d(X).
\]
Recall that $\mathcal{C}^\perp = \langle h_{\mathcal{C}}^*(X)/h_{\mathcal{C}}(0) \rangle$. The reciprocal of a product is the product of the reciprocals. Each $Q_d(X)$ with $d > 1$ is self-reciprocal, and $Q_1^*(X) = -Q_1(X)$. Hence $h_{\mathcal{C}}^*(X) = \varepsilon\, h_{\mathcal{C}}(X)$ with $\varepsilon \in \{1, -1\}$. Comparing leading coefficients (the leading coefficient of $h_{\mathcal{C}}^*$ is $h_{\mathcal{C}}(0)$, and $h_{\mathcal{C}}$ is monic) gives $h_{\mathcal{C}}(0) = \varepsilon$. Therefore $h_{\mathcal{C}}^*(X)/h_{\mathcal{C}}(0) = h_{\mathcal{C}}(X)$, which is monic and divides $X^n - 1$, so it is the generator polynomial $g_{\mathcal{C}^\perp}(X)$.
\end{proof}

\begin{lemma}\label{lem:gerador_potencia_prima}
Let $\ell$ be a prime with $\ell \neq \operatorname{char}(\mathbb{F}_q)$, let $a \ge 1$, and let $\mathcal{C}$ be the cyclic code of length $\ell^a$ with generator polynomial $g_{\mathcal{C}}(X) = Q_{\ell^a}(X)$. Then
\[
g_{\mathcal{C}^\perp}(X) = \frac{X^{\ell^a} - 1}{Q_{\ell^a}(X)} = X^{\ell^{a-1}} - 1 .
\]
\end{lemma}

\begin{proof}
The divisors of $\ell^a$ are $1, \ell, \dots, \ell^a$. By Theorem~\ref{thm:descomposicao_xn-1},
\[
X^{\ell^a} - 1 = \prod_{i=0}^{a} Q_{\ell^i}(X)
\quad \text{and} \quad
X^{\ell^{a-1}} - 1 = \prod_{i=0}^{a-1} Q_{\ell^i}(X).
\]
Hence $\dfrac{X^{\ell^a}-1}{Q_{\ell^a}(X)} = X^{\ell^{a-1}} - 1$. The expression for $g_{\mathcal{C}^\perp}(X)$ follows from Lemma~\ref{lem:dual_ciclotomico} with $S = \{\ell^a\}$.
\end{proof}

	Let $\mathbb{F}_{q}^{\ast} = \mathbb{F}_q \setminus \{ 0 \}$ be the cyclic group of order $q-1$ under the multiplication of $\mathbb{F}_q$ and if $\gamma$ is a generator of this cyclic group, then $\mathbb{F}_ q = \{ 0,1= \gamma^0, \gamma, \ldots, \gamma^{q-2} \}$ and $\gamma^j = 1$ if and only if $q-1$ divides $j$. Let $\mathbb{F}_{q^{\mu}}$ be a finite extension field of $\mathbb{F}_q$. For $\alpha \in \mathbb{F}_{q^{\mu}}$ let $M_{\alpha}(X) \in \mathbb{F}_q[X]$ be the minimal polynomial of $\alpha$ over $\mathbb{F}_q$. The roots of $M_{\alpha}(X)$ are distinct and lie in $\mathbb{F}_{q^{\mu}}$. Since $M_{\alpha}(X^q) = M_{\alpha}(X)^q$, we have that $\alpha, \alpha^q, \alpha^{q^2}, \ldots$ are all roots of $M_{\alpha}(X)$. Note that this sequence stops after $r$ terms, for some integer $r$ where $\alpha^{q^r} = \alpha$. Suppose now that $\gamma$ is a primitive element of $\mathbb{F}_{q^{\mu}}$. Thus, $\alpha = \gamma^s$ for some $s$, and so $\alpha^{q^r} = \alpha$ if and only if $\gamma^{sq^r - s } = 1$. Then, $sq^r \equiv s$ $(\mbox{mod } q^{\mu} -1)$. Based on this, we define the \textit{$q$-cyclotomic coset of $s$ modulo $q^{\mu} - 1$} to be the set $C_s = \{ s, sq, \ldots , sq^{r-1} \}$ $(\mbox{mod } q^{\mu}-1)$, where $r$ is the smallest positive integer such that $sq^r \equiv s$ $(\mbox{mod } q^{\mu} -1)$. The sets $C_s$ partition the set $\{ 0,1,2,\ldots, q^{\mu} -2 \}$ into disjoint sets. We can extend this concept as follows. Let $n$ be a positive integer with $\gcd(n,q)=1$ and let $s$ be an integer with $0 \leq s < n$. The \textit{$q$-cyclotomic coset of $s$ modulo $n$} is the set $C_s = \{ s, sq, \ldots, sq^{r-1} \}$ $(\mbox{mod } n)$, where $r$ is the smallest positive integer such that $sq^r \equiv s$ $(\mbox{mod } n)$.

	Let $\mathcal{C}$ be a cyclic code with generator polynomial $g_{\mathcal{C}}(X) \in R_{q,n}$. Then, by \cite[Theorems 4.1.1 and 4.2.1]{HuffmanPless2003}, $g_{\mathcal{C}}(X) = \prod_{s} \prod_{i \in C_{s}} (X - \alpha^i)$, where $\alpha$ is a primitive $n$-th root of unity (recall that $\gcd(n,q)=1$) and $s$ runs through some subset of representatives of the $q$-cyclotomic cosets $C_s$ modulo $n$. The set $T_{\mathcal{C}}^{\alpha} := \bigcup_{s} C_s$ is called \textit{defining set of} the cyclic code $\mathcal{C}$. Note that, $T_{\mathcal{C}}^{\alpha}  = \{ 0\leq i < n \mbox{ ; } g_{\mathcal{C}}(\alpha^i) = 0\}$.

	We say that $T_{\mathcal{C}}^{\alpha}$ contains $\ell$ consecutive elements provided there is a set $\{ \lambda, \lambda+1, \ldots, \lambda + \ell -1\}$ of $\ell$ consecutive integers such that $\{  \lambda, \lambda+1, \ldots, \lambda + \ell -1 \}$ $(\mbox{mod } n) \subseteq T_{\mathcal{C}}^{\alpha}$.

\begin{theorem}(BCH bound) \cite[Theorem 4.5.3]{HuffmanPless2003}
	\label{thm:cota_bch}
	Let $\mathcal{C}$ be a cyclic code of length $n$ over $\mathbb{F}_q$ with minimum distance $d(\mathcal{C})$ and defining set $T_{\mathcal{C}}^{\alpha}$ . If $T_{\mathcal{C}}^{\alpha}$ contains $\delta - 1$ consecutive elements, then $d(\mathcal{C}) \geq \delta$.
\end{theorem}

\begin{theorem}(Generalized BCH bound) \cite[Theorem 2]{HartmannTzeng1972}
	\label{teo:cota_ht}
	Let $\mathcal{C}$ be a cyclic code of length $n$ over $\mathbb{F}_q$ with minimum distance $d(\mathcal{C})$ and generator polynomial $g_{\mathcal{C}}(X)$. Let $\alpha$ be a primitive $n$-th root of unity. Suppose that there are integers $b, c, \delta$ and $s$ such that $\delta \ge 2$, $s \ge 0$ and $\gcd(c, n) = 1$. If $g_{\mathcal{C}}(\alpha^{b + i + jc})=0$ for all $0 \le i \le \delta - 2$ and $0 \le j \le s$, then
	\begin{equation}
		d(\mathcal{C}) \ge \delta + s.
	\end{equation}
\end{theorem}

For more details about coding theory, see e.g. \cite{LingXing2004}.

Now, assume that $n$ is a composite positive integer and let $t=\operatorname{lpf}(n)$, that is, $t$ is the least prime factor of $n$. We denote by $\mathcal{C}_{n,t}$ and $\mathcal{C}_{n,t,1}$ the cyclic codes in $\mathbb{F}_q^n$ generated, respectively, by the polynomials $Q_n(X) Q_t(X)$ and $Q_n(X)Q_t(X)Q_1(X)$, where $Q_1(X)$, $Q_t(X)$ and $Q_n(X)$ are cyclotomic polynomials as in (\ref{n cyclotomic polynomial}).

In the next sections, we study the cyclic codes $\mathcal{C}_{n,t}$ and $\mathcal{C}_{n,t,1}$ and their respective duals. This study was motivated by two works of A. K. Bhagat and R. Sarma, \cite{BhagatSarma2025, BhagatSarma2026v2}. In those works, Bhagat and Sarma studied the cyclic codes $\mathcal{C}_n$ and $\mathcal{C}_{n,1}$ generated, respectively, by the polynomials $Q_n(X)$ and $Q_n(X)Q_1(X)$, as well as their respective duals, and presented the following results.

	\begin{theorem}\cite[Theorem 3.5]{BhagatSarma2025}
		\label{thm:parametros_cn}
		Let $n > 1$ be an integer with $\gcd(n, p) = 1$. Then $\mathcal{C}_n$ is an $[n, n - \varphi(n), \text{lpf}(n)]_q$ cyclic code.
	\end{theorem}

	\begin{theorem}\cite[Theorem 3.6]{BhagatSarma2025}
		\label{thm:parametros_cn1}
		Let $n$ be a composite integer such that $\gcd(n, p) = 1$. Then $\mathcal{C}_{n,1}$ is an $[n, n - \varphi(n) - 1, 2 \cdot \text{lpf}(n)]_q$ cyclic code.
	\end{theorem}

		\begin{theorem}\cite[Theorem 4.7]{BhagatSarma2026v2} \label{teo:parametros_dual_Cn}
		Let $n > 1$ be an integer with $\gcd(n, p) = 1$. Then $\mathcal{C}_{n}^{\perp}$ is an $[n, \varphi(n), 2^{\omega(n)}]_q$ cyclic code.
			\end{theorem}

	\begin{theorem}\cite[Theorem 4.11]{BhagatSarma2026v2} \label{teo:parametros_dual_Cn1}
		Let $n$ be a composite integer such that $\gcd(n, p) = 1$. Then $\mathcal{C}_{n,1}^{\perp}$ is an $[n,  \varphi(n) + 1, 2^{\omega(n)}]_q$ cyclic code.
		\end{theorem}

\begin{remark} \label{prop:cotas_inferiores_iniciais}
Since $\mathcal{C}_{n,t} \subseteq \mathcal{C}_n$ and $\mathcal{C}_{n,t,1} \subseteq \mathcal{C}_{n,1}$, by Theorems \ref{thm:parametros_cn} and \ref{thm:parametros_cn1}, it follows that
\begin{itemize}
			\item[(i)] $d(\mathcal{C}_{n,t}) \ge t$,
			\item[(ii)] $d(\mathcal{C}_{n,t,1}) \ge 2t$.
		\end{itemize}
\end{remark}

\subsection{LCD and LRC codes}

A \textit{linear complementary dual} (LCD) code is a linear code $\mathcal{C}$
over $\mathbb{F}_q$ such that $\mathcal{C} \cap \mathcal{C}^\perp = \{0\}$. Such codes
were introduced by Massey~\cite{Massey1992}, who showed that they provide an
optimal linear coding solution to the wire-tap channel of type II. More recently,
LCD codes have been shown to provide protection against side-channel attacks and
fault-injection attacks in cryptographic hardware
implementations~\cite{CarletGuilley2016}.

Let $\mathcal{C}$ be a linear $[n, k, d]_q$-code over $\mathbb{F}_q$. The $i$-th
    coordinate of $\mathcal{C}$, with $0 \le i \le n - 1$, is said to have
    \textit{locality} $r$ if the symbol $c_i$ of any codeword
    $c = (c_0, c_1, \dots, c_{n-1}) \in \mathcal{C}$ can be recovered from $r$
    other coordinates, that is, if $c_i$ can be expressed as a function of
    $c_{i_1}, \dots, c_{i_r}$, where $0 \le i_1 < i_2 < \dots < i_r \le n - 1$
    and $i_j \neq i$ for all $j \in \{1, \dots, r\}$.

    The code $\mathcal{C}$ is said to have \textit{locality} $r$ if every coordinate
    has locality at most $r$. In this case, $\mathcal{C}$ is called a
    \textit{locally recoverable code} (LRC) and its parameters are denoted by
    $[n, k, d; r]_q$. Locally recoverable codes were introduced by Gopalan,
    Huang, Simitci, and Yekhanin~\cite{GopalanHuangSimitciYekhanin2012} in the
    context of distributed storage systems, where the need to repair a failed
    storage node by accessing only a small number of surviving nodes motivates the
    locality constraint. A cyclic LRC code that is also LCD is called a
    \textit{cyclic LCD-LRC code}.

The parameters of any LRC code are subject to a generalization of the classical
Singleton bound, known as the Singleton-type bound.

\begin{lemma}[Singleton-type bound { \cite{GopalanHuangSimitciYekhanin2012}, \cite[Lemma 5.2]{BhagatSarma2025}}]
    \label{lema:singleton_lrc}
    Any LRC code over $\mathbb{F}_q$ with parameters $[n, k, d; r]_q$ satisfies
    \begin{equation}
        \label{eq:singleton_lrc}
        d \le n - k - \left\lceil \frac{k}{r} \right\rceil + 2.
    \end{equation}
\end{lemma}

Based on this bound, we adopt the following standard terminology to assess
the quality of the LRC codes constructed:
\begin{itemize}
    \item \textbf{$d$-optimal (or simply optimal):} An LRC code $[n, k, d; r]_q$
    over $\mathbb{F}_q$ that achieves equality in \eqref{eq:singleton_lrc}, that is,
    \[
    d = n - k - \left\lceil \frac{k}{r} \right\rceil + 2.
    \]
    \item \textbf{Almost $d$-optimal:} An LRC code $[n, k, d; r]_q$ over $\mathbb{F}_q$
    such that an LRC code with parameters $[n, k, d+1; r]_q$ would be $d$-optimal
    (i.e., the minimum distance $d$ falls exactly one unit short of the bound
    in \eqref{eq:singleton_lrc}).
    \item \textbf{$r$-optimal:} An LRC code $[n, k, d; r]_q$ over $\mathbb{F}_q$ for
    which no LRC code over $\F_q$ with the same parameters $[n, k, d]_q$
    achieves a strictly smaller locality $s < r$.
\end{itemize}

The following lemma directly relates the locality of a cyclic code to the
minimum distance of its dual, and will be the main tool for determining the
locality of our codes.

\begin{lemma}[{\cite{LXY19}, \cite[Lemma 5.4]{BhagatSarma2025}}]
    \label{lema:localidade_dual}
    Any cyclic code $\mathcal{C}$ over $\mathbb{F}_q$ whose dual distance satisfies
    $d(\mathcal{C}^\perp) \le r + 1$ has locality $r$.
\end{lemma}

\section{The minimum distance of the cyclic codes $\mathcal{C}_{n,t}$ and $\mathcal{C}_{n,t,1}$}
\label{sec:dist_cnt}

Let $q$ be a power of the prime $p$, let $n$ be a composite positive integer coprime to $p$, and let $t=\operatorname{lpf}(n)$. Let $\mathcal{C}_{n,t} = \langle Q_n(X) Q_t(X) \rangle$ and $\mathcal{C}_{n,t,1} = \langle Q_n(X) Q_t(X) Q_1(X) \rangle$ be the cyclic codes in $\mathbb{F}_q^n$ defined in the previous section.

Note that these codes have length $n$ and their dimensions are given by

\begin{equation*}
	\dim(\mathcal{C}_{n,t}) = n - \deg(Q_n(X)Q_t(X)) = n - \varphi(n) - (t - 1) = n - \varphi(n) - t + 1,
\end{equation*}
and
\begin{equation*}
	\dim(\mathcal{C}_{n,t,1}) = n - \deg(Q_n(X)Q_t(X)Q_1(X)) = n - \varphi(n) - (t - 1)-1 = n - \varphi(n) - t.
\end{equation*}

In what follows, we assume that $n > t^2$, which excludes the trivial case for the dimensions of the codes $\mathcal{C}_{n,t}$ and $\mathcal{C}_{n,t,1}$.

\begin{theorem}
	\label{cor:distancia_estagnada}
	Let $n$ and $t=  \operatorname{lpf}(n)$ be as above. If $t^2 \nmid n$, then $d(\mathcal{C}_{n, t}) = t$.
\end{theorem}

\begin{proof}
	By Remark \ref{prop:cotas_inferiores_iniciais}, we have $d(\mathcal{C}_{n, t}) \ge t$. Let $P(X) =  \frac{X^n-1}{X^s-1}$, where $s=n/t$. Then $P(X)  = 1 + X^s + X^{2s} + \dots + X^{(t-1)s}$ has exactly $t$ nonzero terms, since the exponents $0, s, \dots, (t-1)s$ are pairwise distinct and smaller than $n$.

		By Lemma \ref{teo:divisibilidade_P_Qt}, since $t^2 \nmid n$, we have that $Q_t(X)$ divides $P(X)$. Moreover, by Theorem \ref{thm:descomposicao_xn-1}, $Q_n(X)$ divides $X^n-1=P(X)(X^s-1)$. The roots of $X^s-1$ have order dividing $s<n$, while the roots of $Q_n(X)$ have order exactly $n$. Hence $\gcd(Q_n(X),X^s-1)=1$, and thus $Q_n(X)$ divides $P(X)$. Now, $Q_n(X)$ and $Q_t(X)$ are distinct factors of the separable polynomial $X^n-1$, hence coprime, and thus $Q_n(X)Q_t(X)$ divides $P(X)$. Since $\deg P(X) = (t-1)s < n$, $P(X)$ is a nonzero codeword of $\mathcal{C}_{n,t}$ of weight $t$. Therefore, $d(\mathcal{C}_{n,t}) = t$.
	\end{proof}

\begin{theorem}
		\label{teo:distancia_exata_Cnt}
		Let $n$ and $t = \operatorname{lpf}(n)$ be as above. If $t^2 \mid n$, then
		\begin{equation}
			d(\mathcal{C}_{n, t}) = 2t.
		\end{equation}
	\end{theorem}

	\begin{proof}
		Let $\zeta$ be a primitive $n$-th root of unity over $\mathbb{F}_q$. By Equation (\ref{n cyclotomic polynomial}) and Theorem \ref{thm:descomposicao_xn-1}, we conclude that the defining set $T_{\mathcal{C}_{n, t}}^{\zeta} = \{0 \leq i < n \mbox{ ; } Q_{t}(\zeta^i) = 0 \mbox{ or } Q_{n}(\zeta^i) = 0\}$, where $Q_{n}(\zeta^i) = 0$ if and only if $\gcd(n,i) = 1$, and $Q_{t}(\zeta^i) = 0$ if and only if $i=ks$, where $s = \frac{n}{t}$ and $\gcd(k, t) = 1$. Note that $s \in T_{\mathcal{C}_{n, t}}^{\zeta}$, since $k=1$ satisfies the condition above.

		Define the set
		\begin{equation*}
			B := \{ s - (t-1), \dots, s - 1, s + 1, \dots, s + (t-1) \}.
		\end{equation*}

		We claim that $B \subseteq T_{\mathcal{C}_{n, t}}^{\zeta}$. Indeed, let $\ell$ be a prime factor of $n$. If $\ell = t$, since $t^2 \mid n$, we have that $t$ divides $s=\frac{n}{t}$. Thus, $s \pm j \equiv \pm j \pmod t$ and so $t \nmid (s \pm j)$, since $0 < j < t$. Now, let $\ell$ be a prime factor of $n$ with $\ell > t$. Since $n=st$ and $\ell> t$, we have that $\ell \mid s$. Thus, $s \pm j \equiv \pm j \pmod \ell$ and so $\ell \nmid (s \pm j)$, since $0 < j < t < \ell$. Then, we obtain that $\gcd(s \pm j, n) = 1$ for every $j \in \{1, \dots, t-1\}$, and thus $Q_{n}(\zeta^i) = 0$ for all $i \in B$.

		 Therefore, $B \cup \{s\} \subseteq T_{\mathcal{C}_{n, t}}^{\zeta}$. Thus, $T_{\mathcal{C}_{n, t}}^{\zeta}$ contains $2t-1$ consecutive elements and so, by Theorem \ref{thm:cota_bch}, $d(\mathcal{C}_{n, t}) \geq 2t$.

				Now, let $P(X) =  \frac{X^n-1}{X^s-1} = 1 + X^s + X^{2s} + \dots + X^{(t-1)s}$, where $s=\frac{n}{t}$. As in the proof of Theorem \ref{cor:distancia_estagnada}, we have that $Q_{n}(X)$ divides $P(X)$. Using that $Q_{t}(X)$ divides $X^t -1$, we obtain $(X^t - 1)P(X) = \sum_{j=0}^{t-1} X^{js+t} - \sum_{j=0}^{t-1} X^{js}$ belongs to $\mathcal{C}_{n,t}$. Since $n>t^2$, we have $s>t$, so the exponents $js$ and $js+t$, $0 \le j \le t-1$, are pairwise distinct and smaller than $n$. Hence $(X^t - 1)P(X)$ has degree smaller than $n$ and exactly $2t$ nonzero terms, and we have $d(\mathcal{C}_{n, t}) \le 2t$. Therefore, $d(\mathcal{C}_{n, t}) = 2t$.

	\end{proof}

\begin{theorem}
	\label{teo:distancia_exata_qualquer_t}
	Let $n$ and $t =  \operatorname{lpf}(n)$ be as above. Then
	\begin{equation}
		d(\mathcal{C}_{n,t,1}) = 2t.
	\end{equation}
\end{theorem}

\begin{proof}

Let $\zeta$ be a primitive $n$-th root of unity over $\mathbb{F}_q$. By definition, $\zeta$ is a root of $Q_n(X)$. Since $n$ is coprime to $\operatorname{char}(\mathbb{F}_q)$ and $t = \operatorname{lpf}(n)$, Remark \ref{lema:raizes_Qt_primo} gives $Q_t(\zeta^{n/t}) = 0$.
		Moreover, every root of $Q_n(X)$ has the form $\zeta^k$ with $\gcd(k,n)=1$, and $(\zeta^k)^{n/t} = \eta^k$, where $\eta = \zeta^{n/t}$ is a primitive $t$-th root of unity. Since $\gcd(k,t)=1$, $\eta^k$ is also a primitive $t$-th root of unity, so $Q_t\bigl((\zeta^k)^{n/t}\bigr)=0$. As $Q_n(X)$ has simple roots, $Q_n(X)$ divides $Q_t(X^{n/t})$, and so there exists a polynomial $A(X)$ such that $A(X)Q_n(X) = Q_t(X^{n/t})$. Then $A(X) Q_n(X) (X^t-1) = A(X)Q_n(X)Q_t(X)Q_1(X) \in  \mathcal{C}_{n,t,1}$. Now,
	\begin{align}
		A(X) Q_n(X) (X^t-1) &=  Q_t(X^{n/t}) (X^t-1) \nonumber \\
		&= \left( X^{\frac{n(t-1)}{t}} + X^{\frac{n(t-2)}{t}} + \dots + X^{n/t} + 1 \right) (X^t-1) \nonumber \\
		&= \sum_{j=0}^{t-1} X^{\frac{nj}{t} + t} - \sum_{j=0}^{t-1} X^{\frac{nj}{t}}.
	\end{align}

	 Therefore, since $n > t^2$ implies $n/t > t$, the exponents $nj/t + t$ and $nj/t$, $0 \le j \le t-1$, are pairwise distinct and smaller than $n$, and we obtain an element $\negc \in \mathcal{C}_{n,t,1}$ with $\operatorname{wt}(\negc) = 2t$. Hence
	 $d(\mathcal{C}_{n,t,1}) \le 2t$ and the result follows from Remark \ref{prop:cotas_inferiores_iniciais}.

\end{proof}

\section{The minimum distance of $\mathcal{C}_{n,t}^{\perp}$ for $\omega(n) \leq 2$}
\label{sec:dual_omega_le2}

In this section, we study the dual code $\mathcal{C}_{n,t}^{\perp}$. We characterize the generator polynomial of $\mathcal{C}_{n,t}^{\perp}$, determine its exact dimension, and establish an upper bound for its minimum distance, using inclusion properties inherited from the base codes. We use the same notation as in the previous sections: $n$ is a composite integer with $\operatorname{char}(\mathbb{F}_q) \nmid n$ and $t = \operatorname{lpf}(n)$. The assumption $n > t^2$ of Section 2 is not needed here. The first information about $\mathcal{C}_{n,t}^{\perp}$ is presented below.

Since $\mathcal{C}_{n,t} = \langle Q_n(X)Q_t(X) \rangle$ and $\deg(Q_n(X)Q_t(X)) = \varphi(n) + t - 1$, we have $\dim(\mathcal{C}_{n,t}^\perp) = \varphi(n) + t - 1$. Moreover, the parity-check polynomial of $\mathcal{C}_{n,t}$ is $h_{\mathcal{C}_{n,t}}(X) = \frac{X^n - 1}{Q_n(X)Q_t(X)}$, and thus $\mathcal{C}_{n,t}^\perp = \langle h^\perp(X) \rangle$, where $h^\perp(X)$ is the reciprocal polynomial of $h_{\mathcal{C}_{n,t}}(X)$, that is, $h^\perp(X) = X^{\deg(h_{\mathcal{C}_{n,t}})} h_{\mathcal{C}_{n,t}}(X^{-1})$. By Lemma \ref{lem:dual_ciclotomico} (with $S = \{n, t\}$), we have $\mathcal{C}_{n,t}^\perp = \langle h_{\mathcal{C}_{n,t}}(X) \rangle$.

\begin{lemma}
		\label{lema:cota_superior_dual_Cnt}
		Let $n$ and $t$ be as above. Then
		\begin{equation}
			d(\mathcal{C}_{n,t}^\perp) \le 2^{\omega(n)}.
		\end{equation}
	\end{lemma}

	\begin{proof}
	Since $\mathcal{C}_{n,t} \subseteq \mathcal{C}_n$, we have $\mathcal{C}_n^\perp \subseteq \mathcal{C}_{n,t}^\perp$ and thus $d(\mathcal{C}_{n,t}^\perp) \le d(\mathcal{C}_n^\perp)$. So, the result follows from Theorem \ref{teo:parametros_dual_Cn}.

			\end{proof}

\begin{proposition}
		\label{teo:colapso_pa}
		Let $\ell$ be a prime and $n = \ell^a$, where $a \ge 2$ is an integer. Suppose that $\operatorname{char}(\mathbb{F}_q) \nmid n$. Then
		\begin{equation}
			d(\mathcal{C}_{\ell^a, \ell}^\perp) = 2.
		\end{equation}
	\end{proposition}

	\begin{proof}

	Since $n=\ell^a$, we have $\omega(n) = 1$. Thus, by Lemma \ref{lema:cota_superior_dual_Cnt}, it follows that $d(\mathcal{C}_{\ell^a, \ell}^\perp) \le 2$. Note that, for every $a \geq 2$, we have $\deg(Q_{\ell^a}(X)Q_\ell(X)) = \varphi(\ell^a) + \ell - 1 = \ell^a - \ell^{a-1} + \ell - 1 < \ell^a = n$, and hence $\dim(\mathcal{C}_{\ell^a, \ell}) \geq 1$. Now, suppose there exists $\negc \in \mathcal{C}_{\ell^a, \ell}^\perp$ with $\operatorname{wt}(\negc) = 1$. Since $\mathcal{C}_{\ell^a, \ell}^\perp$ is cyclic, it contains all cyclic shifts of $\negc$, and hence all unit vectors. Thus $\mathcal{C}_{\ell^a, \ell}^\perp = \mathbb{F}_q^n$ and $\mathcal{C}_{\ell^a, \ell} = \{\mathbf{0}\}$, which contradicts $\dim(\mathcal{C}_{\ell^a, \ell}) \geq 1$. Therefore, $d(\mathcal{C}_{\ell^a, \ell}^\perp) \ge 2$, and the result follows.

		\end{proof}

		By the previous result, we observe that for $a=2$, that is, $n=\ell^2$, the code $\mathcal{C}_{\ell^2, \ell}^\perp$ has minimum distance equal to $2$, the code $\mathcal{C}_{\ell^2, \ell}$ has parameters $[\ell^2, 1, \ell^2]_q$, and $\mathcal{C}_{\ell^2, \ell}^\perp$ is a $[\ell^2, \ell^2 - 1, 2]_q$-code. In this case, the parameters of both cyclic codes attain the Singleton bound, that is, the codes are MDS.

	\begin{proposition}
		\label{teo:colapso_dual_pq}
		Let $n = p_1p_2$, where $p_1 < p_2$ are distinct prime integers, and $\operatorname{char}(\mathbb{F}_q) \nmid n$. Then
		\begin{equation}
			d(\mathcal{C}_{p_1p_2, p_1}^\perp) = 2.
		\end{equation}

	\end{proposition}

		\begin{proof}
		Note that, $\frac{X^{p_1 p_2}-1}{Q_{p_1p_2}(X)Q_{p_1}(X)} = Q_{p_2}(X) Q_1(X) = X^{p_2}-1$. Thus, $\mathcal{C}_{p_1p_2, p_1}^\perp = \langle X^{p_2} -1 \rangle$ and we get $\negc \in \mathcal{C}_{p_1p_2, p_1}^\perp$ such that $\operatorname{wt}(\negc) =2$. So, $d(\mathcal{C}_{p_1p_2, p_1}^\perp) \leq 2$. Now, suppose that there exists $\negc' \in \mathcal{C}_{p_1p_2, p_1}^\perp$ with $\operatorname{wt}(\negc')=1$. Since $\mathcal{C}_{p_1p_2, p_1}^\perp$ is cyclic, it contains all unit vectors, so $\mathcal{C}_{p_1p_2, p_1}^\perp = \mathbb{F}_q^n$ and $\mathcal{C}_{p_1p_2, p_1} = \{\mathbf{0}\}$. But $\dim(\mathcal{C}_{p_1p_2, p_1}) = n - \varphi(n) - (p_1 - 1) = p_2 > 0$, a contradiction. Therefore, $d(\mathcal{C}_{p_1p_2, p_1}^\perp) = 2$.

	\end{proof}

		\begin{proposition}
		\label{teo:distancia_exata_pa_qb}
		Let  $n = p_{1}^{\lambda_1} p_{2}^{\lambda_2}$, where $p_1 < p_2$ are distinct prime integers, $\lambda_1$ and $\lambda_2$ are positive integers such that  $n > p_1p_2$. If $\operatorname{char}(\mathbb{F}_q) \nmid n$, then
		\begin{equation}
			d(\mathcal{C}_{n, p_1}^\perp) = 4.
		\end{equation}
	\end{proposition}

	\begin{proof}
	Since $\omega(n) = 2$, by Lemma \ref{lema:cota_superior_dual_Cnt}, we have $d(\mathcal{C}_{n, p_1}^\perp)  \leq 4$.

	Let $\alpha$ be a primitive $n$-th root of unity. Let $g^\perp(X) = \frac{X^n - 1}{Q_n(X)Q_{p_1}(X)}$. Since $\mathcal{C}_{n, p_1}^\perp = \langle g^\perp(X)  \rangle$, the roots of $g^\perp(X)$ are not roots of $Q_n(X)Q_{p_1}(X)$. So, for $0 < m < n$, $\alpha^m$ is a root of $g^\perp(X)$ if and only if
	\begin{enumerate}
	\item $\gcd(m,n) \neq 1$, and
	\item by Remark \ref{lema:raizes_Qt_primo}, $m$ is not a multiple of $\dfrac{n}{p_1} = p_{1}^{\lambda_1-1}p_{2}^{\lambda_2}$.
	\end{enumerate}

Now, to apply the bound in Theorem  \ref{teo:cota_ht}, we consider the following systems of congruences:

\begin{align*}
			b &\equiv 0 \pmod{p_1} \\
			b &\equiv p_2-1 \pmod{p_{2}^{\lambda_2}}
		\end{align*}
		and
		\begin{align*}
			c &\equiv -1 \pmod{p_1} \\
			c &\equiv 1 \pmod{p_{2}^{\lambda_2}}.
		\end{align*}

Since $c \equiv -1 \pmod{p_1}$ and $c \equiv 1 \pmod{p_{2}^{\lambda_2}}$, we have that $p_1 \nmid c$ and $p_2 \nmid c$. So, $\gcd(c, n) = 1$. Let $s=1$ and $\delta = 3$ be as in Theorem  \ref{teo:cota_ht}. Thus, we get four possibilities for the exponent $m$ such that $g^\perp(\alpha^m)=0$: $m \in \{b, b+1, b+c, b+c+1\}$.

Note that $p_1 \mid b$ and $p_1 \mid (b+c+1)$. Thus, $\gcd(b,n) \neq 1$ and $\gcd(b+c+1,n) \neq 1$. Moreover, since $b+1 \equiv p_2 \pmod{p_{2}^{\lambda_2}}$ and $b+c \equiv p_2 \pmod{p_2^{\lambda_2}}$, we have $p_2 \mid (b+1)$ and $p_2 \mid (b+c)$. Hence, $\gcd(b+c,n) \neq 1 $ and  $\gcd(b+1,n)\neq 1$. Next, we show that none of $b$, $b+1$, $b+c$ and $b+c+1$ is a multiple of $p_{1}^{\lambda_1-1}p_{2}^{\lambda_2}$.

\begin{itemize}
\item Since $p_2 \mid (b+1)$ and $p_2 \mid (b+c)$, neither $b$ nor $b+c+1$ is divisible by $p_2$. Hence, neither of them is a multiple of $p_1^{\lambda_1-1}p_2^{\lambda_2}$.

\item Since $p_1 \mid b$ and $p_1 \mid (b+c+1)$, neither $b+1$ nor $b+c$ is divisible by $p_1$. If $\lambda_1 \geq 2$, then $p_1 \mid p_1^{\lambda_1-1}p_2^{\lambda_2}$, so neither $b+1$ nor $b+c$ is a multiple of $p_1^{\lambda_1-1}p_2^{\lambda_2}$. If $\lambda_1 = 1$, then $\lambda_2 \geq 2$ and $p_1^{\lambda_1-1}p_2^{\lambda_2} = p_2^{\lambda_2}$. In this case, both $b+1$ and $b+c$ leave remainder $p_2 \neq 0$ modulo $p_2^{\lambda_2}$, so neither is a multiple of $p_1^{\lambda_1-1}p_2^{\lambda_2}$.
\end{itemize}

Thus, if $m \in \{b, b+1, b+c, b+c+1\}$ (taken modulo $n$; none of them is $\equiv 0 \pmod n$, because $0$ is a multiple of $n/p_1$), then $g^\perp(\alpha^m)=0$. Therefore, by Theorem \ref{teo:cota_ht}, $d \ge \delta + s = 3 + 1 = 4$, and the result follows.

\end{proof}

\section{The minimum distance of $\mathcal{C}_{n,t}^{\perp}$ for $\omega(n) \geq 3$}
\label{sec:dual_omega_ge3}

The technique employed to determine the minimum distance of $\mathcal{C}_{n,t}^{\perp}$ in the case $\omega(n) \leq 2$ cannot be extended to the case $\omega(n) \geq 3$. In this case, our approach relies instead on a certain product of codes, which is defined as follows.

\begin{definition}
Given $r \geq 2$ linear codes $\mathcal{C}_1, \dots, \mathcal{C}_r$ over $\mathbb{F}_q$, with parameters $[n_i, k_i, d_i]_q$ for $i = 1, \dots, r$, the \textit{direct product} $\mathcal{C}_1 \otimes \cdots \otimes \mathcal{C}_r$ is defined as the vector space whose codewords can be represented as $r$-dimensional arrays $c = (c_{j_1, \dots, j_r})$, with $j_i \in \mathbb{Z}_{n_i}$, of size $n_1 \times \cdots \times n_r$ over $\mathbb{F}_q$, such that, for each $i \in \{1, \dots, r\}$, every line of $c$ along the $i$-th direction belongs to $\mathcal{C}_i$. That is, for every choice of indices $(j_1, \dots, j_{i-1}, j_{i+1}, \dots, j_r)$, the vector
\[
\bigl(c_{j_1, \dots, j_{i-1}, \, \tau, \, j_{i+1}, \dots, j_r}\bigr)_{\tau=0}^{n_i-1} \in \mathbb{F}_q^{n_i}
\]
is a codeword of $\mathcal{C}_i$.
\end{definition}

In \cite{BhagatSarma2026v2}, Bhagat and Sarma work with the case $r=2$ and observed that if $G_1$ and $G_2$ are generator matrices of the linear codes $\mathcal{C}_1$ and $\mathcal{C}_2$, respectively, then $G_1 \otimes G_2$ is a generator matrix of $\mathcal{C}_1 \otimes \mathcal{C}_2$, where $G_1 \otimes G_2$ denotes the Kronecker product of matrices. Moreover, for two matrices $M_1$ and $M_2$ over $\mathbb{F}_q$, the code generated by $M_1 \otimes M_2$ and the code generated by $M_2 \otimes M_1$ are permutation equivalent. If $\mathcal{C}_1$ and $\mathcal{C}_2$ are both cyclic codes, then $\mathcal{C}_1 \otimes \mathcal{C}_2$ is closed under cyclic shift of all rows simultaneously and cyclic shift of all columns simultaneously. Algebraically, if $\mathcal{C}_1$ is an ideal of $\mathbb{F}_q[X] / \langle X^{n_1} - 1 \rangle$ and $\mathcal{C}_2$ is an ideal of $\mathbb{F}_q[X] / \langle X^{n_2} - 1 \rangle$, and if we identify a typical element $c = (c_{i,j})$ of $\mathcal{C}_1 \otimes \mathcal{C}_2$ by
$$
f(X,Y) = \sum_{i=0}^{n_1-1} \sum_{j=0}^{n_2-1} c_{i,j}X^i Y^j,
$$
then $X f(X,Y)$ and $Y f(X,Y)$ represent the cyclic shift of all rows simultaneously and of all columns simultaneously, respectively. Now, set $\mathcal{R}_{n_1, n_2}:= \mathbb{F}_{q}[X,Y] / \langle X^{n_1} -1, Y^{n_2}-1 \rangle$ and assume $\gcd(n_1, n_2)=1$. By the Chinese Remainder Theorem,  for each $(i,j) \in \mathbb{Z}_{n_1} \times \mathbb{Z}_{n_2}$, there is a unique $\psi(i,j) \in \mathbb{Z}_{n_1 n_2}$ such that
		\begin{align*}
			\psi(i,j) &\equiv i \pmod{n_1}, \\
			\psi(i,j) &\equiv j \pmod{n_2}.
		\end{align*}

		Therefore, we have a ring isomorphism $\Phi: \mathcal{R}_{n_1,n_2} \to \mathbb{F}_q[Z]/\langle Z^{n_1 n_2}-1 \rangle$ given by $\Phi(X^i Y^j) = Z^{\psi(i,j)}$, where $Z = XY$. Thus, for any codeword $f(X,Y) \in \mathcal{C}_1 \otimes \mathcal{C}_2$, we have $g(Z):= \Phi(f(X,Y)) \in \Phi(\mathcal{C}_1 \otimes \mathcal{C}_2)$. Then, Bhagat and Sarma provided the following result.

	\begin{theorem} \cite[Theorem 4.6]{BhagatSarma2026v2}
		\label{teo:permutacao_crt}
		Let $n_1, n_2 > 1$ be integers coprime to $\operatorname{char}(\mathbb{F}_q)$. If $\gcd(n_1, n_2) = 1$, then $\mathcal{C}_{n_1 n_2}^\perp$ and $\mathcal{C}_{n_1}^\perp \otimes \mathcal{C}_{n_2}^\perp$ are permutation equivalent.
	\end{theorem}

In what follows, we generalize this result to an arbitrary number of factors. We first fix some notation.

Let $n_1, \dots, n_j$ be pairwise coprime positive integers. By the Chinese Remainder Theorem, there is a bijection
\[
\Psi_j \colon \mathbb{Z}_{n_1} \times \cdots \times \mathbb{Z}_{n_j} \to \mathbb{Z}_{n_1 \cdots n_j}
\]
that sends $(i_1, \dots, i_j)$ to the unique element $x \in \mathbb{Z}_{n_1 \cdots n_j}$ satisfying
\[
x \equiv i_m \pmod{n_m} \quad \text{for all } m \in \{1, \dots, j\}.
\]

\begin{lemma}\label{lema:composicao_crt}
Let $n_1, \dots, n_k$ be pairwise coprime positive integers and set $n' = \prod_{m=1}^{k-1} n_m$. Let $\Psi_{k-1}$ and $\Psi_k$ be the bijections as above associated with $n_1, \dots, n_{k-1}$ and $n_1, \dots, n_k$, respectively, and let $\psi \colon \mathbb{Z}_{n'} \times \mathbb{Z}_{n_k} \to \mathbb{Z}_{n' n_k}$ be the bijection associated with the coprimes $n'$ and $n_k$. Then, for all $(i_1, \dots, i_k) \in \mathbb{Z}_{n_1} \times \cdots \times \mathbb{Z}_{n_k}$,
\[
\psi\bigl(\Psi_{k-1}(i_1, \dots, i_{k-1}),\, i_k\bigr) = \Psi_k(i_1, \dots, i_k).
\]
\end{lemma}

\begin{proof}
Fix $(i_1, \dots, i_k)$ and set $a = \Psi_{k-1}(i_1, \dots, i_{k-1}) \in \mathbb{Z}_{n'}$ and $x = \psi(a, i_k) \in \mathbb{Z}_{n' n_k}$. By the definition of $\Psi_{k-1}$,
\begin{equation}\label{eq:crt_hyp}
a \equiv i_m \pmod{n_m}, \qquad m = 1, \dots, k-1.
\end{equation}
By the definition of $\psi$,
\begin{align}
x &\equiv a \pmod{n'}, \label{eq:crt_comp1} \\
x &\equiv i_k \pmod{n_k}. \label{eq:crt_comp2}
\end{align}
Since $n_m \mid n'$ for every $m \in \{1, \dots, k-1\}$, congruence \eqref{eq:crt_comp1} implies $x \equiv a \pmod{n_m}$. Combining this with \eqref{eq:crt_hyp} yields
\begin{equation}
x \equiv a \equiv i_m \pmod{n_m}, \qquad m = 1, \dots, k-1.
\end{equation}
Together with \eqref{eq:crt_comp2}, this shows that $x$ satisfies $x \equiv i_m \pmod{n_m}$ for all $m \in \{1, \dots, k\}$. By the Chinese Remainder Theorem, the system of these $k$ congruences has a unique solution modulo $n_1 \cdots n_k = n' n_k$, and by definition of the bijection this solution is $\Psi_k(i_1, \dots, i_k)$. Hence $x = \Psi_k(i_1, \dots, i_k)$, and the result follows.
\end{proof}

This explicit description of the permutation allows us to extend the result, by induction on the number of prime factors, to an arbitrary number of coprime factors.

Let $\mathbb{F}_q^{\,n_1 \times \cdots \times n_j}$ denote the $\mathbb{F}_q$-vector space of arrays
$c = (c_{i_1, \dots, i_j})$ with entries in $\mathbb{F}_q$ indexed by
$\mathbb{Z}_{n_1} \times \cdots \times \mathbb{Z}_{n_j}$. It is isomorphic to the tensor product of the underlying vector spaces:
\[
\mathbb{F}_q^{\,n_1} \otimes \cdots \otimes \mathbb{F}_q^{\,n_j} \;\cong\; \mathbb{F}_q^{\,n_1 \times \cdots \times n_j},
\]
where the isomorphism sends $v_1 \otimes \cdots \otimes v_j$, with $v_m = (v_{m,i})_{i \in \mathbb{Z}_{n_m}}$, to the array $\bigl(v_{1,i_1} v_{2,i_2} \cdots v_{j,i_j}\bigr)_{(i_1, \dots, i_j)}$.

\begin{example}
This example only illustrates the notation; no coprimality of the sizes is needed here. For $j = 2$, an array is a matrix; for $j = 3$, it is a three-dimensional block of entries of $\mathbb{F}_q$, that is, a cube (more precisely, a rectangular parallelepiped) with $n_1$ rows, $n_2$ columns and $n_3$ layers, where $c_{i_1, i_2, i_3}$ is the entry at position $(i_1, i_2, i_3)$.

Let $q = 2$ and $(n_1, n_2, n_3) = (2, 3, 2)$. Thus, an array $c = (c_{i_1, i_2, i_3}) \in \mathbb{F}_2^{\,2 \times 3 \times 2}$ is a $2 \times 3 \times 2$ parallelepiped with $2 \cdot 3 \cdot 2 = 12$ entries. We can visualize it as two layers, obtained by fixing $i_3 = 0$ and $i_3 = 1$, each being a $2 \times 3$ matrix whose rows are indexed by $i_1$ and whose columns are indexed by $i_2$:
\[
c_{\cdot,\cdot,0} = \begin{pmatrix} 1 & 0 & 1 \\ 0 & 1 & 1 \end{pmatrix},
\qquad
c_{\cdot,\cdot,1} = \begin{pmatrix} 0 & 0 & 1 \\ 1 & 1 & 0 \end{pmatrix}.
\]
For instance, $c_{0,0,0} = 1$, $c_{1,2,0} = 1$, $c_{0,2,1} = 1$ and $c_{1,0,1} = 1$.

The lines of $c$ along each direction are the vectors obtained by letting one index vary while the other two are fixed:
\begin{itemize}
    \item along direction~1, for fixed $(i_2, i_3)$, we obtain vectors in $\mathbb{F}_2^{2}$; for example, $(i_2, i_3) = (0, 0)$ gives $(c_{0,0,0}, c_{1,0,0}) = (1, 0)$;
    \item along direction~2, for fixed $(i_1, i_3)$, we obtain vectors in $\mathbb{F}_2^{3}$; for example, $(i_1, i_3) = (0, 0)$ gives $(c_{0,0,0}, c_{0,1,0}, c_{0,2,0}) = (1, 0, 1)$;
    \item along direction~3, for fixed $(i_1, i_2)$, we obtain vectors in $\mathbb{F}_2^{2}$; for example, $(i_1, i_2) = (0, 0)$ gives $(c_{0,0,0}, c_{0,0,1}) = (1, 0)$.
\end{itemize}
The array $c$ belongs to $\mathcal{C}_1 \otimes \mathcal{C}_2 \otimes \mathcal{C}_3$ if and only if, for each $i \in \{1, 2, 3\}$, all lines of $c$ along direction~$i$ are codewords of $\mathcal{C}_i$. The array above is only an arbitrary element of $\mathbb{F}_2^{\,2 \times 3 \times 2}$ and need not satisfy this condition.

Arrays of the form $c_{i_1, i_2, i_3} = v_{1,i_1}\, v_{2,i_2}\, v_{3,i_3}$ correspond to the pure tensors $v_1 \otimes v_2 \otimes v_3$. For example, for $v_1 = (1, 1)$, $v_2 = (1, 0, 1)$ and $v_3 = (1, 0)$, the layer $i_3 = 0$ is $\begin{pmatrix} 1 & 0 & 1 \\ 1 & 0 & 1 \end{pmatrix}$ and the layer $i_3 = 1$ is the zero matrix.
\end{example}

Now, let $n_1, \dots, n_j$ be pairwise coprime positive integers, and let $\mathcal{D} \subseteq \mathbb{F}_q^{\,n_1 \times \cdots \times n_j}$ be a linear code whose coordinates are indexed by $\mathbb{Z}_{n_1} \times \cdots \times \mathbb{Z}_{n_j}$ (ordered, say, lexicographically, so that permutation equivalence makes sense). For $c = (c_{i_1, \dots, i_j}) \in \mathcal{D}$, let $\Psi_j(c) \in \mathbb{F}_q^{\,n_1 \cdots n_j}$ be the word with coordinates indexed by $\mathbb{Z}_{n_1 \cdots n_j}$ given by
\[
\Psi_j(c)_{\Psi_j(i_1, \dots, i_j)} = c_{i_1, \dots, i_j}.
\]
We write $\Psi_j(\mathcal{D}) = \{\Psi_j(c) : c \in \mathcal{D}\}$. Thus $\Psi_j(\mathcal{D})$ is obtained from $\mathcal{D}$ by permuting coordinates, and in particular $\mathcal{D}$ and $\Psi_j(\mathcal{D})$ are permutation equivalent. The same convention applies to $\psi$.

The following refinement of Theorem~\ref{teo:permutacao_crt} shows that the permutation can be taken to be $\psi$. We include a proof, since this is exactly what is used in the proof of Proposition~\ref{prop:decomposicao_tensorial_geral}.

\begin{lemma}\label{lem:crt_dual_explicit}
Let $n_1, n_2 > 1$ be coprime integers, both coprime to $\operatorname{char}(\mathbb{F}_q)$. Then
\[
\mathcal{C}_{n_1 n_2}^\perp = \psi\bigl(\mathcal{C}_{n_1}^\perp \otimes \mathcal{C}_{n_2}^\perp\bigr).
\]
\end{lemma}

\begin{proof}
Put $n = n_1 n_2$ and let $\zeta$ be a primitive $n$-th root of unity in an extension of $\mathbb{F}_q$. Then $\beta_1 = \zeta^{n_2}$ and $\beta_2 = \zeta^{n_1}$ are primitive $n_1$-th and $n_2$-th roots of unity, respectively. For $i = 1, 2$ let $Z_i = \{a \in \mathbb{Z}_{n_i} : \gcd(a, n_i) \neq 1\}$, and let $Z = \{m \in \mathbb{Z}_n : \gcd(m, n) \neq 1\}$. By Lemma~\ref{lem:dual_ciclotomico} with $S = \{n_i\}$, the code $\mathcal{C}_{n_i}^\perp$ is generated by the separable polynomial $\prod_{d \mid n_i,\, d \neq n_i} Q_d(X)$, whose roots are exactly the elements $\beta_i^{a}$ with $a \in Z_i$. Hence a word $u \in \mathbb{F}_q^{n_i}$ lies in $\mathcal{C}_{n_i}^\perp$ if and only if $\sum_{l} u_l \beta_i^{al} = 0$ for all $a \in Z_i$. In the same way, a word $u \in \mathbb{F}_q^{n}$ lies in $\mathcal{C}_{n}^\perp$ if and only if $\sum_{l} u_l \zeta^{ml} = 0$ for all $m \in Z$.

For an array $c = (c_{i,j})$ indexed by $\mathbb{Z}_{n_1} \times \mathbb{Z}_{n_2}$ put $F(a,b) = \sum_{i,j} c_{i,j}\, \beta_1^{ai} \beta_2^{bj}$. We claim that $c \in \mathcal{C}_{n_1}^\perp \otimes \mathcal{C}_{n_2}^\perp$ if and only if $F(a,b) = 0$ whenever $a \in Z_1$ or $b \in Z_2$. Indeed, if $c$ belongs to the tensor product and $a \in Z_1$, then $F(a,b) = \sum_j \beta_2^{bj} \sum_i c_{i,j} \beta_1^{ai} = 0$, because every column of $c$ lies in $\mathcal{C}_{n_1}^\perp$; the case $b \in Z_2$ is symmetric, using the rows. Conversely, suppose that $F$ vanishes as stated, and fix $a \in Z_1$. Then $\sum_j \bigl(\sum_i c_{i,j}\beta_1^{ai}\bigr) \beta_2^{bj} = 0$ for every $b \in \mathbb{Z}_{n_2}$. Since $n_2$ is invertible in $\mathbb{F}_q$, the inverse discrete Fourier transform in $j$ gives $\sum_i c_{i,j}\beta_1^{ai} = 0$ for every $j$. Hence every column of $c$ lies in $\mathcal{C}_{n_1}^\perp$. Symmetrically, every row lies in $\mathcal{C}_{n_2}^\perp$.

Now let $c' = \psi(c)$, that is, $c'_{\psi(i,j)} = c_{i,j}$. Let $u_1, u_2$ be integers with $u_1 n_2 \equiv 1 \pmod{n_1}$ and $u_2 n_1 \equiv 1 \pmod{n_2}$. Then $\psi(i,j) \equiv u_1 n_2\, i + u_2 n_1\, j \pmod{n}$, since both sides are congruent to $i$ modulo $n_1$ and to $j$ modulo $n_2$. Hence, for every $m \in \mathbb{Z}_n$,
\[
\sum_{k} c'_k \zeta^{mk} = \sum_{i,j} c_{i,j}\, \zeta^{m(u_1 n_2 i + u_2 n_1 j)} = F(a, b), \qquad \text{where } a \equiv m u_1 \pmod{n_1},\ b \equiv m u_2 \pmod{n_2}.
\]
Since $u_1$ is invertible modulo $n_1$ and $u_2$ is invertible modulo $n_2$, we have $a \in Z_1$ if and only if $\gcd(m, n_1) \neq 1$, and $b \in Z_2$ if and only if $\gcd(m, n_2) \neq 1$. As $\gcd(n_1, n_2) = 1$, the condition ``$\gcd(m,n_1) \neq 1$ or $\gcd(m,n_2) \neq 1$'' is equivalent to $\gcd(m, n) \neq 1$, that is, to $m \in Z$. Therefore $c \in \mathcal{C}_{n_1}^\perp \otimes \mathcal{C}_{n_2}^\perp$ if and only if $\sum_k c'_k \zeta^{mk} = 0$ for all $m \in Z$, that is, if and only if $c' \in \mathcal{C}_n^\perp$.
\end{proof}

\begin{proposition}
\label{prop:decomposicao_tensorial_geral}
Let $n = p_1^{a_1} p_2^{a_2} \cdots p_k^{a_k}$ be the prime factorization of $n$, where $p_1, \dots, p_k$ are distinct primes, $a_i \ge 1$, $k = \omega(n) \ge 1$, and $\operatorname{char}(\mathbb{F}_q) \nmid n$. Then the base dual code $\mathcal{C}_n^\perp = \langle \frac{X^n-1}{Q_n(X)} \rangle$ is permutation equivalent, via the permutation $\Psi_k$ given by the Chinese Remainder Theorem, to the tensor product of the dual codes associated with the prime powers $p_i^{a_i}$:
\begin{equation}
\mathcal{C}_n^\perp = \Psi_k \left( \mathcal{C}_{p_1^{a_1}}^\perp \otimes \mathcal{C}_{p_2^{a_2}}^\perp \otimes \cdots \otimes \mathcal{C}_{p_k^{a_k}}^\perp \right).
\end{equation}
\end{proposition}

\begin{proof}
We proceed by induction on $k = \omega(n)$. For $k = 1$ the claim is trivial, since $\Psi_1$ is the identity. (For $k = 2$ the claim is exactly Lemma~\ref{lem:crt_dual_explicit}, with $\Psi_2 = \psi$.)

Let $k \ge 2$ and assume the claim holds for integers with $k-1$ distinct prime factors. Set $n' = \prod_{i=1}^{k-1} p_i^{a_i}$ and $n_k = p_k^{a_k}$, so that $n = n' n_k$ and $\gcd(n', n_k) = 1$. Write
\[
\mathcal{D} = \mathcal{C}_{p_1^{a_1}}^\perp \otimes \cdots \otimes \mathcal{C}_{p_{k-1}^{a_{k-1}}}^\perp
\quad \text{and} \quad
\mathcal{E} = \mathcal{C}_{n_k}^\perp.
\]
By the induction hypothesis, $\mathcal{C}_{n'}^\perp = \Psi_{k-1}(\mathcal{D})$. Applying Lemma~\ref{lem:crt_dual_explicit} to the coprime integers $n'$ and $n_k$, we obtain
\begin{equation}\label{eq:prop_step}
\mathcal{C}_n^\perp = \psi\bigl(\mathcal{C}_{n'}^\perp \otimes \mathcal{E}\bigr) = \psi\bigl(\Psi_{k-1}(\mathcal{D}) \otimes \mathcal{E}\bigr).
\end{equation}

It remains to show that $\psi\bigl(\Psi_{k-1}(\mathcal{D}) \otimes \mathcal{E}\bigr) = \Psi_k(\mathcal{D} \otimes \mathcal{E})$. By associativity of the tensor product, $\mathcal{D} \otimes \mathcal{E}$ is the code of arrays $c = (c_{i_1, \dots, i_k})$ whose lines in each direction $m$ belong to the corresponding component code. Equivalently, it is spanned by the arrays $d \otimes e$ with $d \in \mathcal{D}$ and $e \in \mathcal{E}$, where $(d \otimes e)_{i_1, \dots, i_k} = d_{i_1, \dots, i_{k-1}}\, e_{i_k}$. For such an array, Lemma~\ref{lema:composicao_crt} gives
\[
\psi\bigl(\Psi_{k-1}(d) \otimes e\bigr)_{\psi(\Psi_{k-1}(i_1, \dots, i_{k-1}),\, i_k)} = d_{i_1, \dots, i_{k-1}}\, e_{i_k} = \Psi_k(d \otimes e)_{\Psi_k(i_1, \dots, i_k)},
\]
because $\psi(\Psi_{k-1}(i_1, \dots, i_{k-1}), i_k) = \Psi_k(i_1, \dots, i_k)$. Hence the two words coincide on every coordinate. By linearity, $\psi(\Psi_{k-1}(\mathcal{D}) \otimes \mathcal{E}) = \Psi_k(\mathcal{D} \otimes \mathcal{E})$. Substituting into \eqref{eq:prop_step} yields
\[
\mathcal{C}_n^\perp = \Psi_k\bigl(\mathcal{D} \otimes \mathcal{E}\bigr) = \Psi_k \left( \mathcal{C}_{p_1^{a_1}}^\perp \otimes \cdots \otimes \mathcal{C}_{p_k^{a_k}}^\perp \right),
\]
which completes the induction.
\end{proof}

\begin{lemma}
\label{lem:paridade_linhas}
Let $n = p_1^{a_1} \cdots p_k^{a_k}$ be as in Proposition~\ref{prop:decomposicao_tensorial_geral}, let $c \in \mathcal{C}_n^\perp$, and let $\Gamma = \Psi_k^{-1}(c)$ be the array defined by $\Gamma_{i_1, \dots, i_k} = c_{\Psi_k(i_1, \dots, i_k)}$. Fix a direction $j \in \{1, \dots, k\}$, set $L_j = p_j^{a_j - 1}$, and let $u = (u_0, \dots, u_{p_j L_j - 1})$ be a line of $\Gamma$ along direction~$j$. Then
\begin{equation}\label{eq:paridade_macroblocos}
\sum_{m=0}^{p_j - 1} u_{r + mL_j} = u_r + u_{r+L_j} + \cdots + u_{r+(p_j-1)L_j} = 0 \qquad \text{for every } r \in \{0, 1, \dots, L_j-1\}.
\end{equation}
\end{lemma}

\begin{proof}
By Proposition~\ref{prop:decomposicao_tensorial_geral}, $\Gamma \in \mathcal{C}_{p_1^{a_1}}^\perp \otimes \cdots \otimes \mathcal{C}_{p_k^{a_k}}^\perp$. By the definition of the direct product of codes, every line of $\Gamma$ along direction~$j$ (that is, the vector obtained by letting $i_j$ vary while the other coordinates are fixed) is a codeword of $\mathcal{C}_{p_j^{a_j}}^\perp$. Write $L = L_j$, so that the length of $u$ is $p_j L$. By Lemma~\ref{lem:gerador_potencia_prima}, the code $\mathcal{C}_{p_j^{a_j}}^\perp$ is generated by $X^L - 1$. Identifying $u$ with the polynomial $u(X) = \sum_{i=0}^{p_j L - 1} u_i X^i$, we have
\[
u(X) \equiv a(X)\,(X^L - 1) \pmod{X^{p_j L} - 1}
\]
for some polynomial $a(X)$. Since $X^L - 1$ divides $X^{p_j L} - 1$, it follows that
\begin{equation}\label{eq:u_div}
u(X) \equiv 0 \pmod{X^L - 1}.
\end{equation}

Now split $u$ into $p_j$ consecutive blocks of length $L$ (the \emph{macro-blocks}):
\[
u = \big[\, \underbrace{u_0, \dots, u_{L-1}}_{\text{block } 0} \;\big|\; \underbrace{u_L, \dots, u_{2L-1}}_{\text{block } 1} \;\big|\; \cdots \;\big|\; \underbrace{u_{(p_j-1)L}, \dots, u_{p_j L - 1}}_{\text{block } p_j - 1} \,\big].
\]
Equivalently, $u(X) = \sum_{m=0}^{p_j - 1} U_m(X)\, X^{mL}$, where $U_m(X) = \sum_{r=0}^{L-1} u_{r+mL}\, X^r$ has degree less than $L$. Since $X^L \equiv 1 \pmod{X^L - 1}$, we obtain
\[
u(X) \equiv \sum_{m=0}^{p_j - 1} U_m(X) \pmod{X^L - 1},
\]
and the right-hand side has degree less than $L$. By \eqref{eq:u_div}, it is the zero polynomial. Comparing coefficients gives \eqref{eq:paridade_macroblocos}.
\end{proof}

\begin{remark}
Equation~\eqref{eq:paridade_macroblocos} says that, in every line of $\Gamma$ along direction~$j$, the coordinates in positions congruent to $r$ modulo $L_j$ sum to zero. In particular, if $u_r \neq 0$, then $u_{r+mL_j} \neq 0$ for some $m \neq 0$. Thus the nonzero entries of $u$ cannot be isolated within a single macro-block: they are coupled across the $p_j$ macro-blocks. This long-range dependence contrasts with the structure of the periodic subspace $\mathcal{V}_t$ with $t = p_1$, which is defined below.
\end{remark}

To determine the minimum distance of $\mathcal{C}_{n,t}^\perp$ in the case $\omega(n) \ge 3$, we express it as a direct sum of two simpler subspaces. To this end, we introduce the subspace $\mathcal{V}_t \subseteq \mathbb{F}_q^n$ and describe its structure.

From now on, $n$ is a composite integer with $\operatorname{char}(\mathbb{F}_q) \nmid n$, and $t = \operatorname{lpf}(n)$. In particular, $t$ is prime and $t < n$. We identify $\mathbb{F}_q^n$ with $R_{q,n} = \mathbb{F}_q[X]/\langle X^n - 1\rangle$. By Theorem~\ref{thm:descomposicao_xn-1}, $X^n - 1 = \prod_{d \mid n} Q_d(X)$. The factors $Q_d$ are pairwise coprime, because $X^n-1$ is separable (as $\operatorname{char}(\mathbb{F}_q)\nmid n$) and the $Q_d$ have pairwise disjoint sets of roots.

\begin{lemma}
\label{lem:gerador_Vt}
The dual of the cyclic code of length $n$ generated by $Q_t(X)$ is generated by
\[
g_{\mathcal{V}_t}(X) = \frac{X^n - 1}{Q_t(X)} = (X - 1)\bigl(1 + X^t + X^{2t} + \cdots + X^{n-t}\bigr).
\]
\end{lemma}

\begin{proof}
The first equality is Lemma~\ref{lem:dual_ciclotomico} with $S = \{t\}$. Since $t \mid n$, we may write
\[
\frac{X^n - 1}{Q_t(X)} = \frac{X^n - 1}{X^t - 1} \cdot \frac{X^t - 1}{Q_t(X)}.
\]
The first factor is $1 + X^t + X^{2t} + \cdots + X^{n-t}$. Since $t$ is prime, Theorem~\ref{thm:descomposicao_xn-1} gives $X^t - 1 = Q_1(X)\,Q_t(X)$, so the second factor equals $Q_1(X) = X - 1$.
\end{proof}

\begin{definition}
\label{def:Vt}
Let $\mathcal{V}_t \subseteq \mathbb{F}_q^n$ be the cyclic code of length $n$ over $\mathbb{F}_q$ with generator polynomial $g_{\mathcal{V}_t}(X) = \frac{X^n - 1}{Q_t(X)}$; equivalently, $\mathcal{V}_t = \langle Q_t(X) \rangle^\perp$ (Lemma~\ref{lem:gerador_Vt}).
\end{definition}

\begin{lemma}
\label{lem:estrutura_Vt}
We have $\dim \mathcal{V}_t = t - 1$. Moreover, $\mathcal{V}_t$ consists exactly of the vectors $c \in \mathbb{F}_q^n$ of the form
\[
c(X) = b(X)\bigl(1 + X^t + X^{2t} + \cdots + X^{n-t}\bigr), \qquad \deg b \le t - 1,\ b(1) = 0.
\]
That is, the vectors of $\mathcal{V}_t$ are those obtained by repeating $n/t$ times a block $(b_0, \dots, b_{t-1})$ of length $t$ whose entries sum to zero. In particular, every $c \in \mathcal{V}_t$ satisfies $c_m = c_{m \bmod t}$ for all $m \in \mathbb{Z}_n$.
\end{lemma}

\begin{proof}
By Lemma~\ref{lem:gerador_Vt}, $\deg g_{\mathcal{V}_t} = n - t + 1$, so $\dim \mathcal{V}_t = n - \deg g_{\mathcal{V}_t} = t - 1$. Every codeword is $c(X) = a(X)\, g_{\mathcal{V}_t}(X)$ with $\deg a \le t - 2$, and then $\deg c \le n-1$, so no reduction modulo $X^n - 1$ is needed. Put $b(X) = a(X)(X - 1)$. As $a$ ranges over the polynomials of degree at most $t-2$, $b$ ranges over the polynomials of degree at most $t-1$ with $b(1) = 0$. Then
\[
c(X) = b(X) + X^t b(X) + X^{2t} b(X) + \cdots + X^{n-t} b(X).
\]
Since $\deg b < t$, the shifted copies $X^{jt} b(X)$ occupy disjoint blocks of $t$ consecutive positions. This gives the stated description.
\end{proof}

Having described the codewords of $\mathcal{V}_t$, we now show that $\mathcal{C}_{n,t}^\perp$ is the direct sum of $\mathcal{C}_n^\perp$ and $\mathcal{V}_t$.

\begin{lemma}
\label{lema:soma_direta_dual}
Let $n$ be a composite integer with $\operatorname{char}(\mathbb{F}_q) \nmid n$ and $t = \operatorname{lpf}(n)$. Then
\begin{equation}
\mathcal{C}_{n,t}^\perp = \mathcal{C}_n^\perp \oplus \mathcal{V}_t.
\end{equation}
\end{lemma}

\begin{proof}
Recall that $\mathcal{C}_{n,t} = \langle Q_n(X) Q_t(X) \rangle$. By Lemma~\ref{lem:dual_ciclotomico}, $\mathcal{C}_{n,t}^\perp = \langle g^\perp \rangle$ with $g^\perp(X) = \frac{X^n - 1}{Q_n(X)\,Q_t(X)}$. Write
\[
g_1(X) = \frac{X^n - 1}{Q_n(X)} = \prod_{\substack{d \mid n \\ d \neq n}} Q_d(X),
\qquad
g_2(X) = \frac{X^n - 1}{Q_t(X)} = \prod_{\substack{d \mid n \\ d \neq t}} Q_d(X),
\]
so that $\mathcal{C}_n^\perp = \langle g_1 \rangle$ and $\mathcal{V}_t = \langle g_2 \rangle$ in $R_{q,n}$. Since $t \mid n$ and $t < n$, the divisors $t$ and $n$ are distinct.

\emph{The sum.} The sum of two ideals of $R_{q,n}$ is generated by the gcd of their generators. By the pairwise coprimality of the $Q_d$,
\[
\gcd(g_1, g_2) = \prod_{\substack{d \mid n \\ d \neq n,\, t}} Q_d(X) = g^\perp(X).
\]
Hence $\mathcal{C}_n^\perp + \mathcal{V}_t = \langle g^\perp \rangle = \mathcal{C}_{n,t}^\perp$.

\emph{The sum is direct.} The intersection of two ideals of $R_{q,n}$ is generated by the lcm of their generators. Here $Q_t$ divides $g_1$ and $Q_n$ divides $g_2$, and every other $Q_d$ divides both, so
\[
\operatorname{lcm}(g_1, g_2) = \prod_{d \mid n} Q_d(X) = X^n - 1.
\]
Since $X^n - 1 = 0$ in $R_{q,n}$, we get $\mathcal{C}_n^\perp \cap \mathcal{V}_t = \{\mathbf{0}\}$. Therefore $\mathcal{C}_{n,t}^\perp = \mathcal{C}_n^\perp \oplus \mathcal{V}_t$.
\end{proof}

We now describe how the periodic structure of $\mathcal{V}_t$ looks in the multidimensional picture given by the permutation $\Psi_k$.

\begin{remark}
\label{obs:estrutura_hipermatricial_Vt}
Let $n = p_1^{a_1} \cdots p_k^{a_k}$ with $p_1 < \cdots < p_k$, so that $t = \operatorname{lpf}(n) = p_1$. Let $v \in \mathcal{V}_t$ be a nonzero vector, and let $V = \Psi_k^{-1}(v)$ be the corresponding array, that is, $V_{i_1, \dots, i_k} = v_{\Psi_k(i_1, \dots, i_k)}$.

By Lemma~\ref{lem:estrutura_Vt}, $v_m = v_{m \bmod p_1}$ for all $m \in \mathbb{Z}_n$. By the definition of $\Psi_k$, the index $m = \Psi_k(i_1, \dots, i_k)$ satisfies $m \equiv i_j \pmod{p_j^{a_j}}$ for $j = 1, \dots, k$. Since $p_1 \mid p_1^{a_1}$, it follows that $m \equiv i_1 \pmod{p_1}$. Consequently,
\begin{equation}
V_{i_1, i_2, \dots, i_k} = v_m = v_{m \bmod p_1} = v_{i_1 \bmod p_1}.
\end{equation}
This identity has two consequences:
\begin{enumerate}
\item \textbf{Constancy transversal to direction~$1$.} The entry $V_{i_1, \dots, i_k}$ does not depend on $i_2, \dots, i_k$. Hence every line of $V$ along a direction $j \ge 2$ is constant.
\item \textbf{Periodicity along direction~$1$.} Every line of $V$ along direction~$1$ is obtained by repeating the base block $(v_0, \dots, v_{p_1 - 1})$, whose entries sum to zero by Lemma~\ref{lem:estrutura_Vt}. Since the length of this line is $p_1^{a_1}$, the block is repeated $p_1^{a_1 - 1}$ times.
\end{enumerate}
\end{remark}

\begin{theorem}
\label{teo:distancia_exata_cnt_dual_geral}
Let $n$ be a composite integer with $\operatorname{char}(\mathbb{F}_q) \nmid n$ and $t = \operatorname{lpf}(n)$. If $k = \omega(n) \ge 3$, then
\begin{equation}
d(\mathcal{C}_{n,t}^\perp) = 2^{\omega(n)}.
\end{equation}
\end{theorem}

\begin{proof}
Write $n = p_1^{a_1} \cdots p_k^{a_k}$ with $p_1 < \cdots < p_k$, so that $t = p_1$. Since $k \ge 3$, we have $n \ge p_1 p_2 p_3 > t^2$.

By Lemma~\ref{lema:cota_superior_dual_Cnt}, $d(\mathcal{C}_{n,t}^\perp) \le 2^{\omega(n)}$. It therefore suffices to show that every nonzero $w \in \mathcal{C}_{n,t}^\perp$ satisfies $\operatorname{wt}(w) \ge 2^{k}$.

By Lemma~\ref{lema:soma_direta_dual}, $w = c + v$ with $c \in \mathcal{C}_n^\perp$ and $v \in \mathcal{V}_t$. Let $\Gamma = \Psi_k^{-1}(c)$, $V = \Psi_k^{-1}(v)$ and $W = \Psi_k^{-1}(w) = \Gamma + V$ be the corresponding arrays. Since $\Psi_k$ only permutes coordinates, $\operatorname{wt}(w) = \operatorname{wt}(W)$. By Proposition~\ref{prop:decomposicao_tensorial_geral}, $\Gamma \in \mathcal{C}_{p_1^{a_1}}^\perp \otimes \cdots \otimes \mathcal{C}_{p_k^{a_k}}^\perp$.

\medskip
\noindent\textbf{Case 1: $v = 0$.} Then $w = c$ is a nonzero codeword of $\mathcal{C}_n^\perp$, and $\operatorname{wt}(w) \ge d(\mathcal{C}_n^\perp) = 2^{\omega(n)}$ by Theorem~\ref{teo:parametros_dual_Cn}.

\medskip
\noindent\textbf{Case 2: $v \neq 0$.} (This includes the case $c = 0$.) We use two facts.

\emph{(a) Parity along direction $2$.} Let $L = p_2^{a_2 - 1}$. Every line $u = (u_0, \dots, u_{p_2 L - 1})$ of $\Gamma$ along direction~$2$ is a codeword of $\mathcal{C}_{p_2^{a_2}}^\perp$, which is generated by $X^L - 1$ (Lemma~\ref{lem:gerador_potencia_prima}). Lemma~\ref{lem:paridade_linhas} with $j=2$ gives
\begin{equation}\label{eq:parity_dir2}
\sum_{m=0}^{p_2 - 1} u_{r + mL} = 0 \qquad \text{for every } r \in \{0, \dots, L-1\}.
\end{equation}

\emph{(b) Structure of $V$.} By Remark~\ref{obs:estrutura_hipermatricial_Vt}, $V_{i_1, \dots, i_k} = f(i_1)$, where $f(i_1) = v_{i_1 \bmod p_1}$ and $(v_0, \dots, v_{p_1 - 1})$ is the base block of $v$, whose entries sum to zero (Lemma~\ref{lem:estrutura_Vt}). Since $v \neq 0$, this block is nonzero. A nonzero block with zero sum has at least two nonzero entries, because a single nonzero entry cannot sum to zero. Hence the set
\[
A = \{ i_1 \in \mathbb{Z}_{p_1^{a_1}} : f(i_1) \neq 0 \}
\]
satisfies $|A| \ge 2 p_1^{a_1 - 1}$.

\medskip
 Fix $i_1 \in A$ and $(i_3, \dots, i_k)$, and let $K = f(i_1) \neq 0$. Consider the line $\ell$ of $W$ along direction~$2$ through these coordinates. Its entries are $W_{i_1, i_2, i_3, \dots, i_k} = u_{i_2} + K$, where $u$ is the corresponding line of $\Gamma$. For each $r \in \{0, \dots, L-1\}$, equation \eqref{eq:parity_dir2} gives
\[
\sum_{m=0}^{p_2 - 1} \bigl(u_{r + mL} + K\bigr) = 0 + p_2 K \neq 0,
\]
because $\operatorname{char}(\mathbb{F}_q) \nmid n$ implies $p_2 \neq 0$ in $\mathbb{F}_q$. Hence, for each $r$, at least one of the $p_2$ entries $W$ indexed by $r, r+L, \dots, r+(p_2-1)L$ is nonzero. These $L$ residue classes are disjoint, so $\ell$ contains at least $L = p_2^{a_2 - 1}$ nonzero entries of $W$.

\medskip
 The lines along direction~$2$ indexed by distinct tuples $(i_1, i_3, \dots, i_k)$ are pairwise disjoint. There are
\[
|A| \cdot \prod_{m=3}^{k} p_m^{a_m} \;\ge\; 2\, p_1^{a_1 - 1} \prod_{m=3}^{k} p_m^{a_m}
\]
such tuples with $i_1 \in A$. Therefore
\begin{equation}\label{eq:wt_lower}
\operatorname{wt}(w) \;\ge\; 2\, p_1^{a_1 - 1}\, p_2^{a_2 - 1} \prod_{m=3}^{k} p_m^{a_m} \;=\; \frac{2n}{p_1 p_2}.
\end{equation}

\medskip
 Since $p_1 \ge 2$ and the primes are increasing, $p_m \ge 5$ for every $m \ge 3$. Hence
\[
\frac{n}{p_1 p_2} = p_1^{a_1 - 1}\, p_2^{a_2 - 1} \prod_{m=3}^{k} p_m^{a_m} \;\ge\; \prod_{m=3}^{k} p_m \;\ge\; 5^{k-2}.
\]
Because $k \ge 3$, we have $(5/2)^{k-2} \ge 5/2 > 2$, that is, $2 \cdot 5^{k-2} > 2^k$. By \eqref{eq:wt_lower},
\[
\operatorname{wt}(w) \ge 2 \cdot 5^{k-2} > 2^{k} = 2^{\omega(n)}.
\]

\medskip
In both cases $\operatorname{wt}(w) \ge 2^{\omega(n)}$ for every nonzero $w \in \mathcal{C}_{n,t}^\perp$, so $d(\mathcal{C}_{n,t}^\perp) \ge 2^{\omega(n)}$. Combined with the upper bound, $d(\mathcal{C}_{n,t}^\perp) = 2^{\omega(n)}$.
\end{proof}

	\section{LCD and LRC Codes from $\mathcal{C}_{n,t}$ and $\mathcal{C}_{n,t,1}$}
\label{sec:lcd_lrc}

In this section, we apply the results obtained in the preceding sections to study
the properties of the cyclic codes $\mathcal{C}_{n,t}$ and $\mathcal{C}_{n,t,1}$,
together with their respective duals $\mathcal{C}_{n,t}^\perp$ and
$\mathcal{C}_{n,t,1}^\perp$, in the context of linear complementary dual (LCD) codes
and locally recoverable codes (LRC).

\subsection{LCD Codes Generated by Cyclotomic Polynomials}

We begin by recalling the definition of an LCD code and the sufficient criterion
for cyclic codes used in \cite{BhagatSarma2025}.

\begin{theorem}[{\cite{YM94}, \cite[Theorem 2.8]{BhagatSarma2025}}]
    \label{teo:criterio_lcd}
    Let $\mathcal{C}$ be a cyclic code of length $n$ over $\mathbb{F}_q$ with
    $\charac(\mathbb{F}_q) \nmid n$, generated by $g_{\mathcal{C}}(X)$. Then
    $\mathcal{C}$ is an LCD code if the inverse of every root of $g_{\mathcal{C}}(X)$
    is again a root of $g_{\mathcal{C}}(X)$.
\end{theorem}

As a direct consequence of Theorem~\ref{teo:criterio_lcd} and
Lemma~\ref{lem:dual_ciclotomico}, we show that all codes studied in this paper
are LCD codes.

\begin{theorem}
    \label{teo:codigos_lcd}
    Let $n$ be a composite integer with $\charac(\mathbb{F}_q) \nmid n$ and let
    $t = \lpf(n)$. Then the cyclic codes $\mathcal{C}_{n,t}$, $\mathcal{C}_{n,t,1}$
    and their respective duals $\mathcal{C}_{n,t}^\perp$ and $\mathcal{C}_{n,t,1}^\perp$
    are LCD codes.
\end{theorem}

\begin{proof}
    For any divisor $d \mid n$, the set of roots of $Q_d(X)$ --- that is, the
    primitive $d$-th roots of unity --- is closed under inversion
    $\alpha \mapsto \alpha^{-1}$ (note that for $d = 1$, the unique root of
    $Q_1(X) = X - 1$ is $1 = 1^{-1}$). Consequently, if a polynomial $g(X)$ is
    a product of cyclotomic polynomials, then the inverse of each root of $g(X)$
    is again a root of $g(X)$.

    By definition, the generator polynomials of $\mathcal{C}_{n,t}$ and
    $\mathcal{C}_{n,t,1}$ are $Q_n(X)Q_t(X)$ and $Q_n(X)Q_t(X)Q_1(X)$,
    respectively. Moreover, by Lemma~\ref{lem:dual_ciclotomico}
    (with $S = \{n, t\}$ and $S = \{n, t, 1\}$), the generator polynomials of
    $\mathcal{C}_{n,t}^\perp$ and $\mathcal{C}_{n,t,1}^\perp$ are
    \[
    g_{\mathcal{C}_{n,t}^\perp}(X) = \prod_{\substack{d \mid n \\ d \notin \{n,t\}}}
    Q_d(X)
    \quad \text{and} \quad
    g_{\mathcal{C}_{n,t,1}^\perp}(X) =
    \prod_{\substack{d \mid n \\ d \notin \{n,t,1\}}} Q_d(X).
    \]
    Since all four generator polynomials are products of cyclotomic polynomials,
    the result follows immediately from Theorem~\ref{teo:criterio_lcd}.
\end{proof}


Combining the parameters determined in Sections~2, 3, and~4 with
Theorem~\ref{teo:codigos_lcd} and Lemma~\ref{lema:localidade_dual}, we establish
the locality of the codes $\mathcal{C}_{n,t}$ and $\mathcal{C}_{n,t}^\perp$, as
well as the optimality of several subfamilies with respect to the Singleton-type
bound (Lemma~\ref{lema:singleton_lrc}).

\begin{theorem}
    \label{teo:lrc_geral}
    Let $n = p_1^{a_1} p_2^{a_2} \cdots p_k^{a_k}$ be a composite integer with
    $\charac(\mathbb{F}_q) \nmid n$, where $p_1 < p_2 < \dots < p_k$ are distinct primes,
    $a_i \ge 1$, $k = \omega(n) \ge 2$, and $n > p_1 p_2$. Let $t = \lpf(n) = p_1$.
    \begin{enumerate}
        \item If $t^2 \nmid n$, then:
        \begin{enumerate}
            \item $\mathcal{C}_{n,t}$ is a cyclic LCD-LRC code over $\mathbb{F}_q$
            with parameters
            \[
            [n, \, n - \varphi(n) - t + 1, \, t;\, 2^{\omega(n)} - 1]_q;
            \]
            \item $\mathcal{C}_{n,t}^\perp$ is a cyclic LCD-LRC code over $\mathbb{F}_q$
            with parameters
            \[
            [n, \, \varphi(n) + t - 1, \, 2^{\omega(n)};\, t - 1]_q.
            \]
        \end{enumerate}
        \item If $t^2 \mid n$, then:
        \begin{enumerate}
            \item $\mathcal{C}_{n,t}$ is a cyclic LCD-LRC code over $\mathbb{F}_q$
            with parameters
            \[
            [n, \, n - \varphi(n) - t + 1, \, 2t;\, 2^{\omega(n)} - 1]_q;
            \]
            \item $\mathcal{C}_{n,t}^\perp$ is a cyclic LCD-LRC code over $\mathbb{F}_q$
            with parameters
            \[
            [n, \, \varphi(n) + t - 1, \, 2^{\omega(n)};\, 2t - 1]_q.
            \]
        \end{enumerate}
    \end{enumerate}
\end{theorem}

\begin{proof}
    By Theorem~\ref{teo:codigos_lcd}, the codes $\mathcal{C}_{n,t}$ and
    $\mathcal{C}_{n,t}^\perp$ are cyclic LCD codes. Their dimensions were determined
    in previous sections, and we have $d(\mathcal{C}_{n,t}^\perp) = 2^{\omega(n)}$
    by Proposition~\ref{teo:distancia_exata_pa_qb} (for $\omega(n) = 2$) and
    Theorem~\ref{teo:distancia_exata_cnt_dual_geral} (for $\omega(n) \ge 3$).
    \begin{enumerate}
        \item If $t^2 \nmid n$, Theorem~\ref{cor:distancia_estagnada} gives
        $d(\mathcal{C}_{n,t}) = t$. Since $d(\mathcal{C}_{n,t}^\perp) = 2^{\omega(n)}$,
        Lemma~\ref{lema:localidade_dual} implies that $\mathcal{C}_{n,t}$ has
        locality $2^{\omega(n)} - 1$. Analogously, since
        $(\mathcal{C}_{n,t}^\perp)^\perp = \mathcal{C}_{n,t}$ has minimum
        distance $t$, Lemma~\ref{lema:localidade_dual} ensures that
        $\mathcal{C}_{n,t}^\perp$ has locality $t - 1$.
        \item If $t^2 \mid n$, Theorem~\ref{teo:distancia_exata_Cnt} gives
        $d(\mathcal{C}_{n,t}) = 2t$. Applying Lemma~\ref{lema:localidade_dual}
        to both $\mathcal{C}_{n,t}$ and $\mathcal{C}_{n,t}^\perp$ yields localities
        $2^{\omega(n)} - 1$ and $2t - 1$, respectively. \qedhere
    \end{enumerate}
\end{proof}

When $n$ is even but not divisible by $4$, we have $t = \lpf(n) = 2$ and
$t^2 \nmid n$, which yields an infinite family of $r$-optimal cyclic LCD-LRC
codes with minimum distance $2^{\omega(n)}$.

\begin{corollary}
    \label{cor:lrc_par}
    Let $n = 2 p_2^{a_2} \cdots p_k^{a_k}$ be an integer with
    $\charac(\mathbb{F}_q) \nmid n$, where $2 < p_2 < \dots < p_k$ are odd primes,
    $a_i \ge 1$ for $2 \le i \le k$, $k = \omega(n) \ge 2$, and $n > 2p_2$.
    Then $\mathcal{C}_{n,2}^\perp$ is an $r$-optimal cyclic LCD-LRC code over
    $\mathbb{F}_q$ with parameters
    \[
    [n, \, \varphi(n) + 1, \, 2^{\omega(n)};\, 1]_q.
    \]
\end{corollary}

\begin{proof}
    Since $t = \lpf(n) = 2$ and $4 \nmid n$, Theorem~\ref{teo:lrc_geral}(1)(b)
    gives that $\mathcal{C}_{n,2}^\perp$ is a cyclic LCD-LRC code with parameters
    $[n, \varphi(n) + 1, 2^{\omega(n)}; 1]_q$. The $r$-optimality of
    $\mathcal{C}_{n,2}^\perp$ is immediate, as $r = 1$ is the smallest possible
    locality.
\end{proof}

We now analyze the case $n = p_1 p_2$, where $p_1 < p_2$ are distinct primes,
obtaining infinite families of cyclic LCD-LRC codes that are simultaneously
$r$-optimal and $d$-optimal.

\begin{theorem}
    \label{teo:lrc_p1p2}
    Let $p_1 < p_2$ be distinct primes and let $n = p_1 p_2$ with
    $\charac(\mathbb{F}_q) \nmid n$. Then:
    \begin{enumerate}
        \item The code $\mathcal{C}_{p_1p_2, p_1}$ is an $r$-optimal cyclic
        LCD-LRC code over $\mathbb{F}_q$ with parameters
        \[
        [p_1 p_2, \, p_2, \, p_1;\, 1]_q.
        \]
        Moreover, if $p_1 = 2$, then $\mathcal{C}_{2p_2, 2}$ is also $d$-optimal,
        with parameters $[2p_2, p_2, 2; 1]_q$.
        \item 
        The dual code $\mathcal{C}_{p_1p_2, p_1}^\perp$ is a cyclic LCD-LRC code over $\mathbb{F}_q$ with parameters   $[p_1 p_2, \, p_2(p_1 - 1), \, 2;\, p_1 - 1]_q$, which is both r-optimal (with locality $r = p_1 - 1$) and d-optimal (with minimum distance $d = 2$).
           \end{enumerate}
\end{theorem}

\begin{proof}
    For $n = p_1 p_2$ with $p_1 < p_2$, we have $t = \lpf(n) = p_1$,
    $n > p_1^2$, and $p_1^2 \nmid n$. Moreover, $\varphi(n) = (p_1 - 1)(p_2 - 1)$,
    so
    \[
    \dim(\mathcal{C}_{p_1p_2, p_1}) = p_1 p_2 - (p_1 - 1)(p_2 - 1) - p_1 + 1
    = p_2
    \]
    and $\dim(\mathcal{C}_{p_1p_2, p_1}^\perp) = p_1 p_2 - p_2 = p_2(p_1 - 1)$.
    By Theorem~\ref{cor:distancia_estagnada} and Proposition~\ref{teo:colapso_dual_pq},
    we have $d(\mathcal{C}_{p_1p_2, p_1}) = p_1$ and
    $d(\mathcal{C}_{p_1p_2, p_1}^\perp) = 2$.
    \begin{enumerate}
        \item Since $d(\mathcal{C}_{p_1p_2, p_1}^\perp) = 2$,
        Lemma~\ref{lema:localidade_dual} implies that $\mathcal{C}_{p_1p_2, p_1}$
        has locality $r = 1$, and hence is $r$-optimal. If $p_1 = 2$, the
        right-hand side of \eqref{eq:singleton_lrc} for $[2p_2, p_2, 2; 1]_q$
        equals
        \[
        2p_2 - p_2 - \left\lceil \frac{p_2}{1} \right\rceil + 2
        = 2 = d(\mathcal{C}_{2p_2, 2}),
        \]
        showing that $\mathcal{C}_{2p_2, 2}$ is $d$-optimal.
        \item Since $d(\mathcal{C}_{p_1p_2, p_1}) = p_1$,
        Lemma~\ref{lema:localidade_dual} ensures that $\mathcal{C}_{p_1p_2,
        p_1}^\perp$ has locality $r = p_1 - 1$. By \eqref{eq:singleton_lrc},
        \[
        d(\mathcal{C}_{p_1p_2, p_1}^\perp) \le p_1 p_2 - p_2(p_1 - 1)
        - \left\lceil \frac{p_2(p_1 - 1)}{p_1 - 1} \right\rceil + 2
        = p_2 - p_2 + 2 = 2,
        \]
        and since $d(\mathcal{C}_{p_1p_2, p_1}^\perp) = 2$, we conclude that
        $\mathcal{C}_{p_1p_2, p_1}^\perp$ is $d$-optimal.

        If $p_1 = 2$, then $r = 1$ and $r$-optimality of $\mathcal{C}_{2p_2,
        2}^\perp$ is immediate. For $p_1 \ge 3$, suppose for contradiction that
        there exists an LRC code over $\mathbb{F}_q$ with parameters
        $[p_1 p_2, p_2(p_1 - 1), 2; s]_q$ for some $1 \le s < p_1 - 1$.
        By \eqref{eq:singleton_lrc}, we would need
        \[
        2 \le p_1 p_2 - p_2(p_1 - 1)
        - \left\lceil \frac{p_2(p_1 - 1)}{s} \right\rceil + 2
        = p_2 - \left\lceil \frac{p_2(p_1 - 1)}{s} \right\rceil + 2,
        \]
        which gives $\left\lceil \frac{p_2(p_1 - 1)}{s} \right\rceil \le p_2$.
        However, since $0 < s < p_1 - 1$, we have $\frac{p_1 - 1}{s} > 1$,
        so $\frac{p_2(p_1 - 1)}{s} > p_2$ and hence
        $\left\lceil \frac{p_2(p_1 - 1)}{s} \right\rceil \ge p_2 + 1$,
        a contradiction. Therefore $\mathcal{C}_{p_1p_2, p_1}^\perp$ is
        $r$-optimal. \qedhere
    \end{enumerate}
\end{proof}


We now consider the case in which the length $n$ is a prime power,
$n = l^a$ with $a \ge 2$, using Proposition~\ref{teo:colapso_pa} and
Theorem~\ref{teo:distancia_exata_Cnt}.

\begin{theorem}
    \label{teo:lrc_potencia_primo}
    Let $l$ be a prime and let $n = l^a$ with $a \ge 2$ and
    $\charac(\mathbb{F}_q) \nmid n$. Then:
    \begin{enumerate}
        \item For $a = 2$, the codes $\mathcal{C}_{l^2, l}$ and
        $\mathcal{C}_{l^2, l}^\perp$ are both $r$-optimal and $d$-optimal cyclic
        LCD-LRC codes over $\mathbb{F}_q$, with parameters
        \[
        [l^2, \, 1, \, l^2;\, 1]_q \quad \text{and} \quad
        [l^2, \, l^2 - 1, \, 2;\, l^2 - 1]_q,
        \]
        respectively.
        \item For $a \ge 3$:
        \begin{enumerate}
            \item $\mathcal{C}_{l^a, l}$ is an $r$-optimal cyclic LCD-LRC code
            over $\mathbb{F}_q$ with parameters
            \[
            [l^a, \, l^{a-1} - l + 1, \, 2l;\, 1]_q.
            \]
            Moreover, if $l = 2$ (with $q$ odd), then $\mathcal{C}_{2^a, 2}$
            is also $d$-optimal for every $a \ge 3$, with parameters
            $[2^a, 2^{a-1} - 1, 4; 1]_q$.
            \item $\mathcal{C}_{l^a, l}^\perp$ is a cyclic LCD-LRC code over
            $\mathbb{F}_q$ with parameters
            \[
            [l^a, \, l^a - l^{a-1} + l - 1, \, 2;\, 2l - 1]_q.
            \]
        \end{enumerate}
    \end{enumerate}
\end{theorem}

\begin{proof}
    By Theorem~\ref{teo:codigos_lcd}, $\mathcal{C}_{l^a, l}$ and
    $\mathcal{C}_{l^a, l}^\perp$ are cyclic LCD codes, and by
    Proposition~\ref{teo:colapso_pa}, $d(\mathcal{C}_{l^a, l}^\perp) = 2$ for
    all $a \ge 2$.
    \begin{enumerate}
        \item For $a = 2$, we noted in Section~3 (immediately after
        Proposition~\ref{teo:colapso_pa}) that $\mathcal{C}_{l^2, l}$ and
        $\mathcal{C}_{l^2, l}^\perp$ have parameters $[l^2, 1, l^2]_q$ and
        $[l^2, l^2 - 1, 2]_q$, respectively. By Lemma~\ref{lema:localidade_dual},
        the locality of $\mathcal{C}_{l^2, l}$ is $r = 2 - 1 = 1$ (hence
        $r$-optimal) and the locality of $\mathcal{C}_{l^2, l}^\perp$ is
        $r = l^2 - 1$.

        For $\mathcal{C}_{l^2, l}$, the right-hand side of \eqref{eq:singleton_lrc}
        is $l^2 - 1 - \lceil 1/1 \rceil + 2 = l^2 = d(\mathcal{C}_{l^2, l})$,
        so $\mathcal{C}_{l^2, l}$ is $d$-optimal. For $\mathcal{C}_{l^2, l}^\perp$,
        the right-hand side of \eqref{eq:singleton_lrc} is
        \[
        l^2 - (l^2 - 1) - \left\lceil \frac{l^2 - 1}{l^2 - 1} \right\rceil + 2
        = 1 - 1 + 2 = 2 = d(\mathcal{C}_{l^2, l}^\perp),
        \]
        so $\mathcal{C}_{l^2, l}^\perp$ is also $d$-optimal. Furthermore, if
        there existed an LRC code with parameters $[l^2, l^2 - 1, 2; s]_q$ for
        some $1 \le s < l^2 - 1$, then $\lceil (l^2 - 1)/s \rceil \ge 2$, and
        \eqref{eq:singleton_lrc} would yield
        $2 \le 1 - \lceil (l^2 - 1)/s \rceil + 2 \le 1$, a contradiction.
        Hence $\mathcal{C}_{l^2, l}^\perp$ is $r$-optimal.

        \item For $a \ge 3$, we have $n = l^a > l^2$ and $l^2 \mid n$.
        By Theorem~\ref{teo:distancia_exata_Cnt}, $d(\mathcal{C}_{l^a, l}) = 2l$.
        Since $\varphi(l^a) = l^a - l^{a-1}$, the dimensions of
        $\mathcal{C}_{l^a, l}$ and $\mathcal{C}_{l^a, l}^\perp$ are
        $l^{a-1} - l + 1$ and $l^a - l^{a-1} + l - 1$, respectively. Applying
        Lemma~\ref{lema:localidade_dual}, we find that $\mathcal{C}_{l^a, l}$
        has locality $r = 2 - 1 = 1$ (hence $r$-optimal) and
        $\mathcal{C}_{l^a, l}^\perp$ has locality $r = 2l - 1$.

        Finally, for $l = 2$, the parameters of $\mathcal{C}_{2^a, 2}$ are
        $[2^a, 2^{a-1} - 1, 4; 1]_q$. Evaluating the right-hand side of
        \eqref{eq:singleton_lrc} gives
        \[
        2^a - (2^{a-1} - 1) - \left\lceil \frac{2^{a-1} - 1}{1} \right\rceil
        + 2 = 2^a - 2(2^{a-1} - 1) + 2 = 4 = d(\mathcal{C}_{2^a, 2}),
        \]
        so $\mathcal{C}_{2^a, 2}$ is $d$-optimal for every $a \ge 3$. \qedhere
    \end{enumerate}
\end{proof}

Finally, we determine the LCD-LRC parameters of the code $\mathcal{C}_{n,t,1}$
and its dual $\mathcal{C}_{n,t,1}^\perp$.

\begin{theorem}
    \label{teo:lrc_Cnt1}
    Let $n$ be a composite integer with $\charac(\mathbb{F}_q) \nmid n$ and let
    $t = \lpf(n)$ with $n > t^2$. Then:
    \begin{enumerate}
        \item The code $\mathcal{C}_{n,t,1}$ is a cyclic LCD-LRC code over $\mathbb{F}_q$
        with parameters
        \[
        [n, \, n - \varphi(n) - t, \, 2t;\, 2^{\omega(n)} - 1]_q.
        \]
        Moreover, if $n = l^a$ for some prime $l$ and $a \ge 3$, or if
        $n = p_1 p_2$ for distinct primes $p_1 < p_2$, then $\mathcal{C}_{n,t,1}$
        is an $r$-optimal cyclic LCD-LRC code with locality $r = 1$, that is,
        with parameters $[l^a, l^{a-1} - l, 2l; 1]_q$ and
        $[p_1 p_2, p_2 - 1, 2p_1; 1]_q$, respectively. In particular, for
        $p_1 = 2$, the code $\mathcal{C}_{2p_2, 2, 1}$ is simultaneously
        $r$-optimal and $d$-optimal, with parameters $[2p_2, p_2 - 1, 4; 1]_q$.
        \item The dual code $\mathcal{C}_{n,t,1}^\perp$ is a cyclic LCD-LRC code
        over $\mathbb{F}_q$ with parameters
        \[
        [n, \, \varphi(n) + t, \, \le 2^{\omega(n)};\, 2t - 1]_q,
        \]
        with $d(\mathcal{C}_{n,t,1}^\perp) = 2$ in the cases $n = l^a$ ($a \ge 3$)
        and $n = p_1 p_2$ ($p_1 < p_2$).
    \end{enumerate}
\end{theorem}

\begin{proof}
    By Theorem~\ref{teo:codigos_lcd}, $\mathcal{C}_{n,t,1}$ and
    $\mathcal{C}_{n,t,1}^\perp$ are cyclic LCD codes. As computed in Section~2,
    $\dim(\mathcal{C}_{n,t,1}) = n - \varphi(n) - t \ge 1$ (since $n > t^2$),
    so $\dim(\mathcal{C}_{n,t,1}^\perp) = \varphi(n) + t < n$ and, by the same
    argument as in Proposition~\ref{teo:colapso_pa}, $d(\mathcal{C}_{n,t,1}^\perp)
    \ge 2$. By Theorem~\ref{teo:distancia_exata_qualquer_t},
    $d(\mathcal{C}_{n,t,1}) = 2t$.

    On the other hand, the inclusions $\mathcal{C}_{n,t,1} \subseteq
    \mathcal{C}_{n,t} \subseteq \mathcal{C}_n$ imply
    $\mathcal{C}_n^\perp \subseteq \mathcal{C}_{n,t}^\perp \subseteq
    \mathcal{C}_{n,t,1}^\perp$. By Theorem~\ref{teo:parametros_dual_Cn},
    it follows that
    \[
    2 \le d(\mathcal{C}_{n,t,1}^\perp) \le d(\mathcal{C}_{n,t}^\perp)
    \le d(\mathcal{C}_n^\perp) = 2^{\omega(n)}.
    \]
    Applying Lemma~\ref{lema:localidade_dual} with the bound
    $d(\mathcal{C}_{n,t,1}^\perp) \le (2^{\omega(n)} - 1) + 1$, we conclude
    that $\mathcal{C}_{n,t,1}$ has locality $2^{\omega(n)} - 1$, and since
    $d(\mathcal{C}_{n,t,1}) = 2t = (2t - 1) + 1$, the dual code
    $\mathcal{C}_{n,t,1}^\perp$ has locality $2t - 1$.

    When $n = l^a$ ($a \ge 3$) or $n = p_1 p_2$ ($p_1 < p_2$),
    Propositions~\ref{teo:colapso_pa} and~\ref{teo:colapso_dual_pq} give
    $d(\mathcal{C}_{n,t}^\perp) = 2$, so $d(\mathcal{C}_{n,t,1}^\perp) = 2$.
    By Lemma~\ref{lema:localidade_dual}, it follows that $\mathcal{C}_{n,t,1}$
    has locality $r = 2 - 1 = 1$, and is therefore $r$-optimal. Finally, for
    $n = 2p_2$ with $p_2 > 2$ prime, the parameters of $\mathcal{C}_{2p_2, 2, 1}$
    are $[2p_2, p_2 - 1, 4; 1]_q$, and the right-hand side of
    \eqref{eq:singleton_lrc} equals
    \[
    2p_2 - (p_2 - 1) - \left\lceil \frac{p_2 - 1}{1} \right\rceil + 2
    = 4 = d(\mathcal{C}_{2p_2, 2, 1}),
    \]
    showing that $\mathcal{C}_{2p_2, 2, 1}$ is also $d$-optimal.
\end{proof}

We close this section with two numerical examples that illustrate the main results of
the paper. The first covers the case in which $n$ is a product of two distinct primes
($t^2 \nmid n$), and the second covers a prime-power length with $t^2 \mid n$.

\begin{example}
\label{ex:n15}
Let $n = 15 = 3 \cdot 5$, so $t = \operatorname{lpf}(15) = 3$ and $\omega(15) = 2$.
Let $q = 2$. Since $t^2 = 9 \nmid 15$, Theorem~\ref{cor:distancia_estagnada} gives
$d(\mathcal{C}_{15,3}) = t = 3$. By Theorem~\ref{teo:distancia_exata_qualquer_t},
$d(\mathcal{C}_{15,3,1}) = 2t = 6$. Since $n = 3 \cdot 5$ is a product of two
distinct primes, Proposition~\ref{teo:colapso_dual_pq} gives $d(\mathcal{C}_{15,3}^\perp) = 2$.
Moreover, $(X^{15}-1)/(Q_{15}(X)Q_3(X)) = Q_1(X)Q_5(X) = X^5 - 1$, so
$\mathcal{C}_{15,3}^\perp = \langle X^5 - 1\rangle$. Using $\varphi(15) = 8$,
the parameters of the four codes are:
\[
\begin{array}{lll}
\mathcal{C}_{15,3} & = \langle Q_{15}(X)Q_3(X)\rangle, & \text{parameters: } [15,\;5,\;3]_2,  \\[4pt]
\mathcal{C}_{15,3,1} & = \langle Q_{15}(X)Q_3(X)Q_1(X)\rangle, & \text{parameters: } [15,\;4,\;6]_2, \\[4pt]
\mathcal{C}_{15,3}^\perp & = \langle X^5 - 1\rangle, & \text{parameters: } [15,\;10,\;2]_2, \\[4pt]
\mathcal{C}_{15,3,1}^\perp, & & \text{parameters: } [15,\;11,\;2]_2.
\end{array}
\]
By Theorem~\ref{teo:codigos_lcd}, all four codes are LCD. By
Theorem~\ref{teo:lrc_p1p2}, $\mathcal{C}_{15,3}$ is an $r$-optimal cyclic LCD-LRC
code with parameters $[15, 5, 3;\, 1]_2$, and $\mathcal{C}_{15,3}^\perp$ is both
$r$-optimal and $d$-optimal with parameters $[15, 10, 2;\, 2]_2$. Indeed,
\[
n - k - \left\lceil \tfrac{k}{r} \right\rceil + 2
= 15 - 10 - \left\lceil \tfrac{10}{2} \right\rceil + 2 = 2 = d(\mathcal{C}_{15,3}^\perp),
\]
so the Singleton-type bound~\eqref{eq:singleton_lrc} is tight. By
Theorem~\ref{teo:lrc_Cnt1}, $\mathcal{C}_{15,3,1}$ is an $r$-optimal cyclic LCD-LRC
code with parameters $[15, 4, 6;\, 1]_2$. For both $\mathcal{C}_{15,3}$ and
$\mathcal{C}_{15,3,1}$, the locality $r = 1$ follows from
$d(\mathcal{C}_{15,3}^\perp) = 2$ via Lemma~\ref{lema:localidade_dual}.
\end{example}

\begin{example}
\label{ex:n8}
Let $n = 8 = 2^3$, so $t = \operatorname{lpf}(8) = 2$ and $\omega(8) = 1$. Let $q$
be an odd prime power with $\gcd(8, q) = 1$ (e.g., $q = 5$). Since $t^2 = 4 \mid 8$
and $n = 8 > 4 = t^2$, Theorem~\ref{teo:distancia_exata_Cnt} gives
$d(\mathcal{C}_{8,2}) = 2t = 4$. By Theorem~\ref{teo:distancia_exata_qualquer_t},
$d(\mathcal{C}_{8,2,1}) = 4$. By Proposition~\ref{teo:colapso_pa} ($n = 2^3$),
$d(\mathcal{C}_{8,2}^\perp) = 2$. Using $\varphi(8) = 4$, the parameters are:
\[
\begin{array}{lll}
\mathcal{C}_{8,2} & = \langle Q_8(X)Q_2(X)\rangle, & \text{parameters: } [8,\;3,\;4]_q,  \\[4pt]
\mathcal{C}_{8,2,1} & = \langle Q_8(X)Q_2(X)Q_1(X)\rangle, & \text{parameters: } [8,\;2,\;4]_q, \\[4pt]
\mathcal{C}_{8,2}^\perp, & & \text{parameters: } [8,\;5,\;2]_q,  \\[4pt]
\mathcal{C}_{8,2,1}^\perp, & & \text{parameters: } [8,\;6,\;2]_q.
\end{array}
\]
By Theorem~\ref{teo:codigos_lcd}, all four codes are LCD. By
Theorem~\ref{teo:lrc_potencia_primo}(2a), with $l = 2$ and $a = 3$,
$\mathcal{C}_{8,2}$ is simultaneously $r$-optimal and $d$-optimal with parameters
$[8, 3, 4;\, 1]_q$. The $d$-optimality is verified directly:
\[
n - k - \left\lceil \tfrac{k}{r} \right\rceil + 2
= 8 - 3 - \left\lceil \tfrac{3}{1} \right\rceil + 2 = 4 = d(\mathcal{C}_{8,2}).
\]
By Theorem~\ref{teo:lrc_potencia_primo}(2b), $\mathcal{C}_{8,2}^\perp$ is a cyclic
LCD-LRC code with parameters $[8, 5, 2;\, 3]_q$, where the locality $2l - 1 = 3$
follows from $d(\mathcal{C}_{8,2}) = 4$ via Lemma~\ref{lema:localidade_dual}.
By Theorem~\ref{teo:lrc_Cnt1} (with $n = 2^3$, $a = 3 \ge 3$), $\mathcal{C}_{8,2,1}$
is an $r$-optimal cyclic LCD-LRC code with parameters $[8, 2, 4;\, 1]_q$,
where the locality $r = 1$ follows from $d(\mathcal{C}_{8,2}^\perp) = 2$ via
Lemma~\ref{lema:localidade_dual}.
\end{example}

\end{document}